\DeclareUnicodeCharacter{2032}{\ensuremath{^{\prime}}}
\documentclass[aps, prd, twocolumn, preprintnumbers, letterpaper, nofootinbib, superscriptaddress, balancelastpage, 10pt]{revtex4-2}
\usepackage[utf8]{inputenc}
\DeclareUnicodeCharacter{2212}{\textminus}
\usepackage{newunicodechar}
\newunicodechar{η}{\eta}
\usepackage{cancel}
\usepackage{textcomp}
\usepackage[utf8]{inputenc}
\usepackage{graphicx}
\usepackage{enumitem}
\usepackage{slashed}
\usepackage{tikz-feynman}
\tikzfeynmanset{compat=1.1.0}
\usepackage{orcidlink}
\usepackage{xcolor}
\usepackage{float}
\usepackage{comment} 
\usepackage{amsmath, amsfonts, amssymb}
\usepackage{bm, bbm}
\usepackage{lipsum}
\usepackage[caption=false]{subfig}
\usepackage{slashed}
\usepackage{tikz-feynman}
\usepackage[normalem]{ulem}
\usepackage{subfloat}
\newcommand{\beq}{\begin{equation}\begin{aligned}{}}
\newcommand{\eeq}{\end{aligned}\end{equation}}
\newcommand{\beqa}[1]{\begin{equation}\begin{aligned}{#1}}
\newcommand{\eeqa}{\end{aligned}\end{equation}}

\newcommand{\bea}{\begin{eqnarray}{}}
\newcommand{\eea}{\end{eqnarray}}

\usepackage{tikz}
\usetikzlibrary{decorations.pathmorphing,arrows.meta,positioning}
\tikzfeynmanset{ with arrow/.style = {
   decoration={
     markings,
     mark=at position 0.5
          with {\arrow[xshift=1mm]{Triangle[black,width=0.8mm,length=1.2mm]}}
     },
   postaction=decorate}
}
\tikzset{
  gluon/.style={decorate, decoration={coil,aspect=0.5,amplitude=2.5pt,segment length=4pt}, thick},
  quark/.style={-{Latex[length=2mm]}, thick},
  axion/.style={dashed, thick},
  vertex/.style={fill=gray!70, draw=black, rectangle, minimum size=4pt, inner sep=0pt},
}
\usetikzlibrary{decorations.pathmorphing}

\tikzset{
	insertion/.pic={kg
		\draw[fill=white, line width=0.8pt] (0,0) circle[radius=0.18];
		\draw[line width=0.8pt] (-0.13,-0.13) -- (0.13,0.13);
		\draw[line width=0.8pt] (-0.13,0.13) -- (0.13,-0.13);
	}
}

\usepackage{hyperref}
\hypersetup{colorlinks=true,citecolor=blue,linkcolor=blue,urlcolor=blue}

\newcommand{\mO}{\mathcal{O}}
\newcommand{\mC}{\mathcal{C}}

\begin{document}

\title{Soft Collinear Effective Theory for Heavy QCD Axions}

\author{Deepanshu Bisht\,\orcidlink{0009-0009-7047-773X}}
\email{dbisht22@iitk.ac.in}
\affiliation{Department of Physics, Indian Institute of Technology, Kanpur-208016, India}

\author{Sabyasachi Chakraborty\,\orcidlink{0000-0001-5356-7607}}
\email{sabyac@iitk.ac.in}
\affiliation{Department of Physics, Indian Institute of Technology, Kanpur-208016, India}

\author{Siddhartha Karmakar\,\orcidlink{0009-0003-0609-9689}}
\email{siddharthak@iitk.ac.in}
\affiliation{Department of Physics, Indian Institute of Technology, Kanpur-208016, India}

\author{Atanu Samanta\,\orcidlink{0009-0006-2782-2911}}
\email{asamanta23@iitk.ac.in}
\affiliation{Department of Physics, Indian Institute of Technology, Kanpur-208016, India}

\begin{abstract}
We develop a soft-collinear effective theory (SCET) framework for heavy QCD axion, considering two low-energy realizations and taking $B\to Ka$ as a benchmark mode. In the first realization, $aG\widetilde G$ is assumed to be the only independent axion interaction at the scale $\mu\sim m_b$. We show that eliminating the redundant flavor-changing derivative-gluon operator in the weak effective theory generates a new dimension-seven axion operator identified as $\mathcal{O}_{\partial ag}$. We match this operator onto SCET and derive the corresponding leading-power soft and spectator-scattering contributions to $B\to Ka$. We obtain a factorized expression for the spectator contribution in terms of perturbative hard kernels and the $B$- and $K$-meson light-cone distribution amplitudes. The spectator contribution arises at the same order in the power expansion as the soft-overlap term and amounts to approximately $25\%$ of the soft contribution. In the second realization, the Wilson coefficient of $aG\widetilde G$ is assumed to be present above the electroweak scale. Renormalization-group evolution and matching then induce a direct $b\to sa$ operator, which subsequently results in a dominant soft form-factor contribution, whereas the gluonic spectator term turns out to be numerically subleading ($\sim 6-7\%$). We thus identify the conditions under which spectator scattering becomes relevant for heavy-axion production in rare $B$-meson decays. Finally, we derive the corresponding bounds on the axion decay constant $f_a$ for both realizations and compare their phenomenological implications.
\end{abstract}

\maketitle
\section{Introduction}
Flavor-changing neutral-current (FCNC) decays of $B$ mesons provide one of the most sensitive probes of physics beyond the Standard Model. In particular, such rare decays involving an axion or axion-like particle (ALP) in the final state have received considerable attention in recent years~\cite{Izaguirre:2016dfi,Bauer:2020jbp,Chakraborty:2021wda,Bertholet:2021hjl,Bauer:2021mvw,Bisht:2024hbs}, as they offer an excellent opportunity to probe weakly coupled pseudoscalars in the mass range $\mathcal{O}(100~\mathrm{MeV})\lesssim m_a\lesssim\mathcal{O}(1~\mathrm{GeV})$. In most phenomenological analyses, the flavor-changing transition is first computed perturbatively and subsequently embedded into the mesonic decay through the $B\to K$ transition form factor, extracted from light-cone sum rules (LCSR). The corresponding scalar form factor is commonly parameterized as~\cite{Ball:2004ye,Ball:2004rg}
\begin{equation}
f_0(q^2)=
\frac{0.33}
{1-q^2(\mathrm{GeV})^2/37.5}\;,
\label{eq:f0}
\end{equation}
and is evaluated at $q^2=m_a^2$ when computing the decay width $\Gamma(B\to Ka)$. Here $f_0$ is the full QCD matrix element of the flavor-changing current. It is, however, often interpreted as a purely soft-overlap quantity,  in which the spectator quark is treated as soft throughout the decay and does not participate in any hard interaction. This identification, however, is only approximate. In the large-recoil factorization~\cite{Beneke:2000wa,Beneke:2003pa} it decomposes as
\begin{equation}
\begin{aligned}
f_0(q^2)&=C_0(E_K)\,\zeta(E_K)+\Delta_{\rm spec}~,\\
~{\rm with }~\Delta_{\rm spec} &=\frac{f_B f_K m_B}{4E_K^2}\,\phi_B^+\!\otimes T\!\otimes\phi_K\;.
\label{eq:f0split}
\end{aligned}
\end{equation}
Here $\zeta$ is the genuinely non-factorizable soft-overlap form factor and $\Delta_{\rm spec}$ is the hard-collinear spectator correction to the current, appearing at $\mathcal{O}(\alpha_s)$. $\phi_B^+$ and $\phi_K$ are the leading-twist $B$ and $K$ meson light-cone distribution amplitudes (LCDAs), and $T$ is the perturbative hard-scattering kernel. Such factorization was first established using QCD for exclusive non-leptonic $B$-decays~\cite{Beneke:2000ry}. Therefore, identifying $f_0$ directly with $\zeta$ overestimates the pure soft-overlap contribution by an amount of order $\Delta_{\rm spec}$. Also, the spectator scattering of the Standard Model (SM) current gets tacitly folded into the soft term. Therefore, it is better to avoid this contamination by using a dedicated determination of the soft form factor. The light-cone quark-model result within soft-collinear effective theory (SCET)~\cite{Lu:2007sg} provides $\zeta$ directly, with the hard-spectator correction to the current not yet included, and gives $\zeta^{B\to K}(m_B/2) = 0.297$.

The conventional parametrization is expressed in terms of the total form factor and therefore does not resolve the underlying momentum regions, obscuring their separate power counting. This becomes particularly important for new-physics operators that contribute predominantly, or even exclusively, through spectator scattering, since such effects are not captured by the conventional form-factor description. A systematic treatment therefore requires an effective theory that separates the soft, collinear, and hard-collinear modes from the outset. In the present work, we employ SCET to achieve this separation in the context of heavy QCD axions.

Throughout this work, we consider the specific scenario in which the only independent axion interaction is
\begin{equation}
  \mathcal{L}_{\rm ALP}=\frac{1}{2}\left(\partial_\mu a\right)^2+\frac{1}{2}m_a^2 a^2+\frac{\mC_{gg}}{f_a}\,a\,G_{\mu\nu}^A\widetilde G^{A\mu\nu}\;,
\label{eq:axionL}
\end{equation}
where $\widetilde G^{A\mu\nu} \equiv (\epsilon^{\mu\nu\rho\sigma}/2) G_{\rho\sigma}^A$, the superscript `$A$' denoting the $SU(3)_c$ adjoint color index. The Wilson coefficient $\mathcal{C}_{gg} = \alpha_s/8\pi$ in most UV realizations of the axion~\cite{Kim:1979if,Shifman:1979if}. It is important to mention that from the perspective of a bottom-up effective field theory, the effective Lagrangian should contain all operators consistent with the symmetries of the theory up to a given mass dimension. Such a complete operator basis, including derivative couplings to quark currents together with electroweak gauge boson operators, is required to ensure renormalizability order by order in the effective field theory expansion~\cite{Chakraborty:2021wda}. Consequently, the minimal axion-gluon interaction should generally be regarded as one part of a larger operator basis rather than the complete low-energy description. In the present work, however, we deliberately isolate the purely gluonic realization in order to understand the hadronic consequences of the fundamental $aG\widetilde G$ interaction itself, without contamination from additional flavor-changing operators. This allows us to identify the spectator-scattering mechanism generated directly by the axion-gluon coupling and to study its phenomenological importance. Also, our consideration is minimal and aligns directly with the well-motivated scenario of heavy QCD axion.

Having established the framework, we first consider the effective theory at $\mu\sim m_b$, assuming that the only non-vanishing Wilson coefficient is associated with the operator $aG\widetilde G$. In this case, no electroweak renormalization-group evolution or electroweak matching enters the analysis. To the best of our knowledge, such a situation has not been considered before in the literature. A central observation of this work is that, through the modified gluon equation of motion in the presence of $aG\widetilde{G}$, the SM operators generate the dimension-seven operator $O_{\partial ag}$ which mediates flavor transitions. After matching to SCET, we show that this operator gives rise to a new leading-power spectator-scattering contribution, which has been absent in the conventional form-factor description and dominates over the contributions induced by the SM four-quark and chromomagnetic operators. To study such an analysis, one naturally requires a framework capable of systematically separating the momentum modes relevant to heavy-to-light decays, namely the heavy $b$-quark, energetic collinear partons inside the kaon, hard-collinear gluons, and the soft spectator quark. SCET provides exactly this. It is well known that SCET emerged originally from such considerations in heavy-to-light flavor-changing $B$ decays~\cite{Bauer:2000ew}. Although the applications of SCET have branched out to diverse fields, it remains a subject of active research in $B$-physics, with recent studies on form factor analyses~\cite{Zhang:2026izv, Bell:2026ktc, Gao:2024vql, Cui:2022zwm} and inclusion of QED effects~\cite{Huang:2023nli, Boer:2023vsg}. See also, Ref.~\cite{Cornella:2026lkp} for a thorough SCET analysis of $B^-\to \mu^- \bar{\nu}_\mu(\gamma)$. SCET furnishes a systematic separation of momentum with manifest power expansion~\cite{Bauer:2000ew, Bauer:2000yr} and a factorized description of soft-overlap and spectator-scattering contributions~\cite{Beneke:2003pa,Beneke_2006,Bauer:2002aj,Bauer:2004tj}. The primary objective of this work is therefore to construct the corresponding SCET factorization directly from the underlying axion-gluon interaction and to determine the leading-power spectator-scattering contribution to the $B\to Ka$ amplitude. 

In the second scenario, we take the more conventional approach where the same Wilson coefficient is taken to be nonzero above the electroweak scale. In this case, renormalization-group evolution and electroweak matching generate a direct flavor-changing operator, $\partial_\mu a\;\bar{s}\gamma^\mu P_L b$, at two-loop order~\cite{Chakraborty:2021wda,Bauer:2021mvw,Bisht:2024hbs}. Its contribution to $B\to Ka$ is conventionally described by the $B\to K$ scalar form factor and enters predominantly through the soft-overlap amplitude, as discussed in Eq.~\eqref{eq:f0}. We analyze this scenario separately and show that the electroweak-generated soft contribution has the same power counting but is numerically much larger than the corresponding spectator-scattering term. Consequently, phenomenological analyses based solely on the total form factor $f_0(q^2)$~\cite{Bisht:2024hbs} remain well justified within their theoretical uncertainties. The relative importance of soft overlap and spectator scattering therefore depends crucially on whether the gluonic axion interaction is introduced below or above the electroweak scale. 

The paper is organized as follows. In Sec.~\ref{sec:sec2}, we construct the axion operator basis below the electroweak scale and derive the dimension-seven operator $\mathcal{O}_{\partial ag}$ using the modified gluon equation of motion. In Sec.~\ref{sec:sec3}, we match $\mathcal{O}_{\partial ag}$ onto SCET and calculate its leading-power spectator-scattering and soft-overlap contributions systematically. In Section~\ref{sec:sec4} analogous contributions from the insertion of $aG\widetilde{G}$ and the chromomagnetic operator are calculated. In Sec.~\ref{sec:sec5}, we study the implications of the direct $b\to sa$ operator generated when $aG\widetilde G$ is present above the electroweak scale, including both its soft and spectator contributions.  Section~\ref{sec:sec6} presents our numerical results pertaining to both cases, discusses the resulting soft-spectator hierarchies, and summarizes the phenomenological implications. The numerical inputs used in our analysis, a brief discussion of ALP-meson mixing effects with SM decays, and comments on the renormalization group evolution effects are collected in the appendices~\ref{app:Numbers}, \ref{app:mixing} and \ref{app:rg}, respectively. We note in passing that we work at leading order in $\alpha_s$ throughout and utilize the position-space formulation of SCET~\cite{Beneke:2002ph}, adopting standard conventions/notations from the literature.

\section{Axion Operator Basis below the electroweak scale}
\label{sec:sec2}
The weak transition $b\to s$ at the scale $\mu\sim m_b$ is described within the framework of Weak Effective Theory (WET), obtained after integrating out the heavy electroweak degrees of freedom~\cite{Buchalla:1995vs}. In the SM, the effective Hamiltonian consists of the familiar current-current, QCD-penguin, electroweak-penguin, and dipole operators. Once axions or axion-like particles are introduced, the low-energy theory consists of
\begin{equation}
\mathcal{L}_{\rm eff}=\mathcal{L}_{\rm WET}+\mathcal{L}_{\rm ALP}\;,
\end{equation}
where $\mathcal{L}_{\rm ALP}$ is given in Eq.~\eqref{eq:axionL}. Within this framework, the leading flavor-changing interactions are generated from the interplay of the SM weak Hamiltonian and the axion-gluon coupling.

The flavor-changing transition $b\to sg$ in the SM arises at one-loop with $W$-boson and top quark propagators. Below the electroweak scale, these are integrated out, and the effects are incorporated in the Wilson coefficients $\mathcal{C}_{Dg}$ and $\mathcal{C}_{8g}$ of the effective dimension-six operators~\cite{Grinstein:1990tj, Simma:1993ky}
\begin{align}
    \mathcal{O}_{Dg}&=\bar s\gamma^\mu P_L T^A b(D^\nu G_{\nu\mu}^A)\;,\nonumber \\
    \mathcal{O}_{8g}&=m_b\;\bar s\sigma^{\mu\nu}P_R T^A b G_{\mu\nu}^A\;.
    \label{eq:WET}
\end{align}
The corresponding Wilson coefficients are 
\begin{equation}
    \mathcal{C}_{Dg}
    =
    i\lambda_t
    \frac{G_F}{\sqrt2}
    \frac{g_s}{4\pi^2}
    E_0\;,
    \quad
    \mathcal{C}_{8g}
    =
    -i\lambda_t
    \frac{G_F}{\sqrt2}
    \frac{g_s}{4\pi^2}
    E_0'\;,
\end{equation}
where $\lambda_t \equiv V_{ts}^*V_{tb}$ and $E_0, E_0'$ contain the Inami-Lim functions $F_1(x_t), F_2(x_t)$ respectively~\cite{Inami:1980fz} where $x_t \equiv m_t^2/M_W^2$. With this, one obtains comparable values for the Wilson coefficients as $|E_0| \simeq |E_0'| \simeq 0.2$. However, at low energies, the full $b\to sg$ vertex also contains contributions from up and charm quarks running in the loop from the insertion of relevant charged-current operators present in the WET basis. Including these effects and using CKM unitarity to replace $V_{cs}^* V_{cb} = - V_{ts}^*V_{tb} - V_{us}^*V_{ub}$ and ignoring the numerically small second term, one finds the effective values: $|E_0| \simeq 5$~\cite{Ahmady:1997fa,Eeg:2005bq, Hou:1997wy}   and $|E_0'| \simeq 0.15$~\cite{Buchalla:1995vs, Bauer:2004tj}, which implies $|\mathcal{C}_{Dg}| \gg |\mathcal{C}_{8g}|$~\footnote{Strictly, $E_0$ is a $q^2$ dependent form factor containing nonlocal effects from charm loop, rather than a local Wilson coefficient. Treating it as part of the coefficient of $O_{Dg}$ holds for the gluon virtualities $ \ll 4m_c^2$~\cite{Khodjamirian:2010vf}. This is satisfied for hard collinear gluons; however, for hard gluons it is only an approximation.}. This hierarchy naively means that the $B\to Ka$ effects descending from $\mathcal{O}_{Dg}$ will have significant numerical dominance over corresponding effects from $\mathcal{O}_{8g}$.

Importantly, however, $\mathcal{O}_{Dg}$ is not retained as an independent operator in the WET basis~\cite{Simma:1993ky}. Instead, one employs the gluon equation of motion,
\begin{equation}
    (D_\mu G^{\alpha\mu})^A
    =
    g_sJ^{\alpha A}\;,
\end{equation}
to rewrite $\mathcal{O}_{Dg}$ in terms of the four-quark penguin operators $\mO_3-\mO_6$ \cite{Grinstein:1990tj}. An important consequence of our construction is that the gluon equation of motion is modified in the presence of the operator $aG\widetilde G$. As a result, eliminating the redundant operator $\mathcal{O}_{Dg}$ no longer produces only the familiar QCD-penguin four-quark operators of the WET basis, but also generates a new dimension-seven ALP operator, as shown below:
\begin{equation}
    (D_\mu G^{\alpha\mu})^A
    =
    g_sJ^{\alpha A}
    +
    \frac{4\mathcal{C}_{gg}}{f_a}
    (\partial_\mu a)
    \widetilde G^{\alpha\mu A}\;.
\end{equation}
Consequently, Eq.~\eqref{eq:WET} now yields
\begin{align}
    \mathcal{O}_{Dg}
    =&\;
    g_s
    (\bar s\gamma^\mu P_L T^A b)
    \sum_q
    (\bar q\gamma_\mu T^A q)
    \nonumber\\
    &
    +
    \frac{4\mathcal{C}_{gg}}{f_a}
    (\bar s\gamma^\mu P_L T^A b)
    (\partial^\nu a)
    \widetilde G_{\mu\nu}^A\;.
\end{align}
The first term reproduces the familiar QCD four-fermi operators of the WET basis, while the second term is a genuinely new dimension-seven interaction induced by the axion-gluon coupling. We therefore define
\begin{equation}
    \mathcal{O}_{\partial ag}
    \equiv
    \bar s\;T^A\gamma^\mu P_L b
    (\partial^\nu a)
    \widetilde G_{\mu\nu}^A\;,
    \label{eq:Odag}
\end{equation}
with the Wilson coefficient 
\begin{equation}
\frac{\mathcal{C}_{\partial ag}}{f_a} =\frac{4\mathcal{C}_{gg}\mathcal{C}_{Dg}}{f_a}\;.
\end{equation}
In the following section, we perform the tree-level matching of this new operator onto SCET, where it gives rise to a spectator-scattering contribution to the $B\to Ka$ amplitude.

\section{Leading-Power SCET Operator and Matching}
\label{sec:sec3}
Integrating out the hard modes at the scale $m_b$ matches the QCD operators onto a basis of SCET operators with definite momentum scaling. This separation makes the power counting manifest and allows the soft-overlap and spectator-scattering contributions to be treated independently within a systematic expansion in the SCET parameter $\lambda$. In the present work, we first concentrate on the operator $O_{\partial ag}$, which provides the dominant contribution to the decay amplitude. The matching of the remaining operators follow analogously and will be discussed later. Throughout, we work at the leading-order matching accuracy; the renormalization-group evolution of the resulting operators below $\mu_h \sim m_b$, and the extent to which it affects the ratio of spectator to soft contributions, is discussed briefly in Appendix~\ref{app:rg}.

\subsection{Spectator contribution from: \texorpdfstring{$\mathcal{O}_{\partial a g}$}{O[∂ag]}} 
\begin{figure}[h]
\centering
\begin{tikzpicture}[line cap=round, line join=round, scale=0.8]
		\coordinate (T)  at (0,0);
		\coordinate (B)  at (0,-1);
		\coordinate (mid)  at (0,-1.5);
		\coordinate (S)  at (0,-3.10);
		\coordinate (Bg) at (-2.5,0); 
		\draw[line width=1.8pt] (-3.6,0) node[above right] {\(h_v (p_b^\mu)\)} -- (0,0);
		\draw[line width=1.8pt] (0,0) -- (3.6,0) node[above left] {\(\xi_n(u p_K^\mu)\)};
		\draw[line width=1.8pt] (-3.6,-3.10) node[above right] {\(q_s(k_s^\mu)\)} -- (0,-3.10);
		\draw[line width=1.8pt] (0,-3.10) -- (3.6,-3.10) node[above left] {$q_n(\bar u p_K^\mu)$};
		
		\draw[
		line width=1.6pt,
		dashed,
		dash pattern=on 9pt off 8pt
		] (T) -- (2.7,1) node[right] {\(a(q^\mu)\)};
		
		\draw[
		line width=1.4pt, purple,
		decorate,
		decoration={coil, aspect=0.65, segment length=7pt, amplitude=5pt}
		] (T) to (S);
        \node at (.8,-1.4) {\(\ell^\mu\)};
		
		\fill[gray] (-0.18,-0.18) rectangle (0.18,0.18)
		node[above, black] {\({\cal O}_{\partial ag}\)} ;
	\end{tikzpicture}
\caption{Leading SCET contribution to $B\to Ka$ mediated by the operator $\mathcal{O}_{\partial a g}$. The gluon in red represents the hard-collinear mode.}
\label{fig:fig1}
\end{figure}
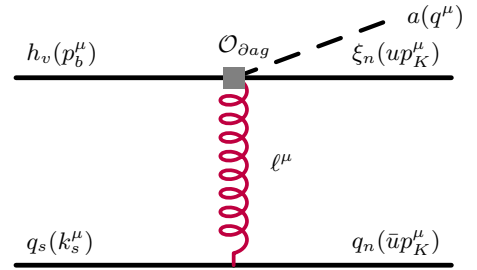

The leading spectator contribution to the form factor is depicted in Fig.~\ref{fig:fig1}. The momentum scaling relevant to the heavy-to-light current with an $n$-hard-collinear gluon is summarized in Table~\ref{tab:scaling-Odag}.
They determine the SCET power counting underlying the matching and factorization analysis. The momentum scaling of external legs is standard: $p_b, p_K, k_s$ denote $b$-quark, kaon, and $B$-meson spectator quark momenta, respectively. The collinear momentum fractions are denoted by $u$ and $\bar{u} = 1-u$, as shown in Fig.~\ref{fig:fig1}. The non-relativistic(NR)/heavy scaling of the axion follows from two-body decay momentum conservation.\footnote{This assignment, however, is not unique. See the discussion at Eq.~\eqref{eq:ndotqaxion}. For $m_a$ comparable to or smaller than $m_K$, the axion becomes $\bar{n}$-collinear with momenta $q \sim (1, \lambda^4, \lambda^2)$.}  The internal gluon momentum becomes $n$-hard collinear (assuming $u, \bar{u} \sim \lambda^0$~\cite{Lange:2003pk}) with the transverse/perp component being non-generic~\cite{Beneke:2003pa}. 
\begin{table}[h]
\centering
\begin{tabular}{|l|l|l|}
\hline
Momentum & Scaling & Mode \\
\hline
\hline
$p_b^\mu=m_b v^\mu+r^\mu$ & $(1,1,\lambda^2)$ & Heavy \\
$k_s^\mu$ & $(\lambda^2,\lambda^2,\lambda^2)$ & Soft \\
$p_K^\mu$ & $(\lambda^4,1,\lambda^2)$ & $n$-collinear \\
$\ell^\mu=k_s^\mu-\bar u\,p_K^\mu$ & $(\lambda^2,1,\lambda^2)$ & $n$-hard-collinear \\
$q^\mu=p_b^\mu + k_s^\mu -p_K^\mu$ & $(1,1,\lambda^2)$ & Heavy/NR axion \\
\hline
\end{tabular}
\caption{Momentum scaling for the spectator contribution induced by the operator $O_{\partial ag}$. The momentum components are ordered as $(n\!\cdot p,\bar n\!\cdot p,p_\perp)$. The SCET power counting parameter is given by $\lambda^2 \sim \Lambda_{\text{QCD}}/m_b$.}
\label{tab:scaling-Odag}
\end{table}

\subsubsection{SCET Operator Construction}
We begin with the field strength tensor $G_{\mu\nu}^A$ of Eq.~\eqref{eq:Odag}. For the spectator-scattering contribution, the gluon field, as shown in Table~\ref{tab:scaling-Odag} must be $n$-hard collinear. For $n$-hard collinear gluon, the leading component of the field strength is
\begin{align}
    G^{A\rho\sigma} = &\frac12 \left(
        n^\rho G_{\bar n\perp}^{A\sigma}-
        n^\sigma G_{\bar n\perp}^{A\rho}
    \right)
    +\mathcal O(\lambda^2)\;, \nonumber \\
    \text{where}\quad 
    & G_{\bar n\perp}^{A\sigma}\equiv\bar n^\alpha G_{\alpha\beta}^A g_\perp^{\beta\sigma}\;,
\end{align}
and $\lambda$ is the SCET power counting parameter. Substituting this into the operator and using the antisymmetry of the Levi-Civita tensor, we find
\begin{equation}
    \mathcal O_{\partial ag} =
    \frac12 \bar s \gamma^\mu P_L T^A b\,
    (\partial^\nu a)\, \epsilon_{\mu\nu\rho\sigma} n^\rho
    G_{\bar n\perp}^{A\sigma}
    +...
\end{equation}
where the ellipses denote terms that are power-suppressed in the SCET expansion.

The operator obtained above is not yet written in terms of the gauge-invariant building blocks of SCET. Gauge invariance under $n$-collinear gauge transformations is restored by introducing the collinear Wilson line $W_n$~\cite{Beneke:2002ph, Bauer:2003mga}, allowing the hard-collinear field strength to be expressed through the standard SCET gluonic building block such as~\cite{Becher:2014oda, bauer_stewart}
\begin{equation}\label{eq:gluonbb}
W_n^\dagger G_{\mu\nu}^A T^A W_n =   \partial_\mu \mathcal{B}_\nu - \partial_\nu \mathcal{B}_\mu - i[\mathcal{B}_\mu, \mathcal{B}_\nu] \equiv \mathcal{G}_{\mu\nu}\;.
\end{equation}
Similarly, the light QCD quark fields are matched onto their corresponding collinear gauge-invariant SCET fields according to 
\begin{equation}\label{eq:colquarkbb}
     \bar s \rightarrow \bar\chi_n\;,\qquad \chi_n=W_n^\dagger\xi_n \;.
\end{equation}
The collinear quark and gluon fields appearing within the building blocks in Eqs.~\eqref{eq:gluonbb}, \eqref{eq:colquarkbb} are still charged under the soft gauge transformations. This is taken care of by performing BPS field redefinition (or \textit{decoupling} transformation) ~\cite{Bauer:2001yt, bauer_stewart, Becher:2014oda} on the collinear quark and gluon fields using the soft Wilson line $Y_s$.
\begin{align}
      \xi_n = Y_s\,\xi_n^{(0)}\,,
  \qquad
  {\cal B}^{A\sigma}_{n\perp}T^A
  = Y_s\, {\cal B}^{(0)A\sigma}_{n\perp}T^A\, Y_s^\dagger \,,
  \label{eq:bps}
\end{align}
where $Y_s$ is the soft Wilson line along the light-like direction $n$~\cite{Bauer:2001yt},
\begin{align}
  Y_s(x) &= P\exp\!\left[\, i g_s\!\int_{-\infty}^{0}\!\!ds\;
  n\cdot A_s(x+sn)\right] .
  \label{eq:Yn}
\end{align}
This removes any soft-collinear interaction from the leading-order SCET Lagrangian. Furthermore, the heavy $b$-quark field appearing in the heavy-to-light current is matched onto its SCET counterpart in the same way as in the heavy-quark effective theory,
\begin{align}
b(x) = e^{-i m_b v\cdot x}\, h_v(x) \,+\, \mathcal{O}(1/m_b)\;.
\label{eq:bmatch}
\end{align}
The redefinition in Eq.\,\eqref{eq:bps} results in the appearance of a $Y_s^\dagger$ inside the SCET operator sitting alongside the $h_v$ field. Note that for notational convenience, we have used $\xi_n$ and ${\cal B}_{n\perp}$ in places of $\xi_n^{(0)}$ and ${\cal B}_{n\perp}^{(0)}$, respectively, throughout the text. At leading power, thus the operator $\mathcal{O}_{\partial a g}$ thus turns out to be
\begin{align}
    \mathcal O_{\partial ag}^{\rm SCET}=
    \frac12\left[\bar\chi_n\gamma_{\perp}^\mu P_L T^A Y_s^\dagger h_v \right]
    (\partial^\nu a)\;
    \epsilon_{\mu\nu\rho\sigma}\; n^\rho \mathcal{G}^{A\sigma}_{\bar{n}\perp}  \;.
    \label{eq:Oag-operator}
\end{align}
Note that here we used the totally antisymmetric property of $\epsilon_{\mu\nu \rho \sigma}$ and the following lightcone decomposition of $\gamma^\mu$ and the SCET projection identity
\begin{equation}
\begin{aligned}
 \gamma^\mu  = \slashed{\bar{n}} \frac{n^\mu}{2} + \slashed{n} \frac{\bar{n}^\mu}{2} + \gamma^\mu_\perp\;, \quad \text{and}\quad \bar{\xi}_n \slashed{n} = 0\;,
\end{aligned}
\end{equation}
after which only the transverse Dirac structure survives. Finally, Eq.~\eqref{eq:Oag-operator} can be further simplified by expanding $\partial^\nu a$ into light cone components as
\begin{equation}
    \partial^\nu a = n^\nu \frac{\bar{n}.\partial a}{2}+\bar{n}^\nu \frac{n.\partial a}{2}+\partial^\nu_\perp a\;,
\end{equation}
and using the antisymmetry properties of the Levi-Civita tensor. Moreover, working at the leading $\alpha_s$ order, the third commutator term within $\mathcal{G}_{\bar{n}\perp}^{A\sigma}$ can be ignored. This simplifies the relevant part of the interaction Lagrangian further
\begin{equation}
\begin{aligned}
    \mathcal{L}^{\rm SCET}_{\partial ag}\supset &\frac{i\mathcal{C}_{\partial a g}}{2f_a}\left[\bar{\chi}_n \gamma_{\perp}^\mu P_L T^A Y_s^\dagger h_v\right]
    \epsilon_{\mu\sigma}^\perp \\
    &(n\cdot\partial a)(\bar{n}\cdot\partial)\mathcal B_{n\perp}^{A\sigma}\;,
    \label{eq:SCETL1}
\end{aligned}
\end{equation}
where the two-dimensional antisymmetric tensor is defined as $\epsilon^\perp_{\mu\sigma}\equiv\epsilon_{\mu\sigma\nu\rho}\;\bar{n}^\nu n^\rho/2\;.$ The subscript `$n(\bar{n})$' on the single Lorentz index object $\mathcal{B}$ denote it being $n(\bar{n})$-hard-collinear unlike the tensorial field strengths $G$ or $\mathcal{G}$ where this subscript denotes one of the Lorentz index contracted with the light-cone vector $n^\mu (\bar{n}^\mu)$. It is worth emphasizing that, up to this stage, no assumptions have been made regarding the kinematic nature of the axion field. In particular, the derivation of the SCET operator is completely independent of whether the axion is relativistic or non-relativistic. Nevertheless, as discussed before Table~\ref{tab:scaling-Odag}, the decay kinematics along with the axion mass determines the momentum scaling $q^\mu$ of the axion. A more quantitative discussion is provided after Eq.~\eqref{eq:nraxion}. 

\subsubsection{Spectator matrix element}\label{subsubsec:spectatorcontribution}
The contribution to the spectator-scattering amplitude for the decay $B\to Ka$ proceeds as follows. The hard-collinear gluon in Eq.~\eqref{eq:SCETL1} interacts with the soft spectator quark via the SCET Lagrangian interaction~\cite{Beneke:2002ph, Bauer:2004tj}
\begin{equation}
\mathcal{L}_{\xi q{\cal B}}^{\rm SCET}=\left(\bar q_sY_s\right)ig_s{\cal B}_{n\perp}^{A\rho}\gamma_{\perp\rho}T^A\left(W_n^\dagger\xi_n\right)\;,
\label{eq:Lxiq}
\end{equation}
thereby generating the leading spectator contribution to the decay amplitude. This is obtained from the time-ordered product of Eqs.~\eqref{eq:SCETL1} and \eqref{eq:Lxiq},
\begin{equation}
\mathcal{M}_{\text{spec}}^{\partial a g} \equiv i\int d^4y\,\langle Ka\left|T\left\{\mathcal{L}_{\partial ag}^{\rm SCET}(0),\mathcal{L}^{\rm SCET}_{\xi q{\cal B}}(y)\right\}\right|B\rangle\;.
\label{eq:Tproduct}
\end{equation}

Introducing the standard mode expansion for the axion field, the one-particle outgoing axion state can be expressed as $a(x)\sim e^{+iq\cdot x}a_v^\dagger(x)$. Therefore, from Eq.~\eqref{eq:SCETL1}, we immediately obtain
\begin{equation}
\langle a(q)|n\cdot\partial a(0)|0\rangle=i\,n\cdot q\;,
\label{eq:nraxion}
\end{equation}
where $q$ is the outgoing momentum for the axion. We also expect the standard factorization for the axion to hold at the $1/f_a$ order. The kinematic regime of the axion is fully determined by the kaon's collinearity (which follows in the large-recoil limit considered here) and the axion's mass. A simple kinematical analysis shows that if we take $2E_K = xm_B$ where $x\lesssim 1$, then
\begin{equation}\label{eq:ndotqaxion}
\begin{aligned}
     n\cdot q &\simeq m_B \bigg[  1 - \frac{m_K^2}{xm_B^2} + \mathcal{O}(\lambda^8) \bigg] \;,\\
     \bar{n}\cdot q &\simeq  m_B \bigg[ 1 - x + \frac{m_K^2}{x m_B^2} + \mathcal{O}(\lambda^8) \bigg] \;,\\
     m_a^2 &= m_B^2\bigg[1 - x + \frac{m_K^2}{m_B^2} \bigg] \;.
\end{aligned}
\end{equation}
For the least possible values of $x$, consistent with the large-recoil/collinear kaon regime, an upper bound on axion mass follows. E.g, $x = 4/5 \Rightarrow m_a \simeq 2.3$ GeV. On the other hand, as $x \to 1$, $m_a \to m_K$, and in this limit, one can simply infer that the axion must become $\bar{n}$-collinear, which has the momentum scaling $q^\mu \sim (1, \lambda^4, \lambda^2)$. However, this does not follow from the generic scaling rules applied to $q^\mu = p_b^\mu + k_s^\mu - p_K^\mu$ since $\bar{n}.q  = \bar{n}.p_b + \bar{n}.k_s - \bar{n}.p_K \sim 1 + \lambda^2 - 1 \sim 1 $ where the fine-tuned cancellation between large $b$-quark and kaon momenta along $n^\mu$ is not captured. Instead, from Eq.~\eqref{eq:ndotqaxion} in the $x\to 1$ limit, we find that the scaling is indeed $\bar{n}$-collinear. Nonetheless, our final results are unaffected whether $q^\mu$ is taken heavy or collinear, so we remain agnostic about $m_a$ and keep the generic $q^\mu \sim (1,1,\lambda^2)$ heavy/NR scaling for concreteness. Finally, kinematically $m_a$ can also become smaller than $m_K$, but this is precluded by the phenomenologically motivated heavy QCD axion mass range.

Substituting Eq.~\eqref{eq:nraxion} into Eq.~\eqref{eq:Tproduct} and from Eq.~\eqref{eq:ndotqaxion}, $n.q = m_B + \mathcal{O}(\lambda^4)$, the spectator-scattering amplitude becomes
\begin{align}
\mathcal{M}_{\rm spec}^{\partial a g}&=\left(g_s m_B\right)\frac{i\mathcal{C}_{\partial ag}}{2f_a}\;\epsilon_{\perp}^{\mu\rho}\int d^4y\int\frac{d^4\ell}{(2\pi)^4}\frac{\bar n\cdot\ell}{\ell^2+i0} e^{i\ell\cdot y}
\nonumber \\
\times
&\langle K|(\bar\xi_{n\alpha} W_n)_i(0)\left(\gamma_{\perp\mu}P_L\right)_{\alpha\beta}(T^A)_{ij}(Y_s^\dagger h_{v\beta})_j\nonumber \\
&(\bar q_{s\delta}Y_s)_k(y)\left(\gamma_{\perp\rho}\right)_{\delta\eta}(T^A)_{k\ell}(W_n^\dagger\xi_{n\eta})_\ell(y)|B\rangle\;,
\label{eq:spec1}
\end{align}
where $\alpha, \beta, \delta, \eta$ are the Dirac spinor indices, $i,j,k,\ell$ are the fundamental $SU(3)_c$ color indices and $A$'s are the adjoint gluon color index. We have also used the expression of the propagator by contracting the two hard-collinear gluon fields carrying momentum $\ell$
\begin{equation}
\langle0|T\left\{\mathcal{B}_{n\perp}^{A\sigma}(0)\mathcal{B}_{n\perp}^{C\rho}(y)\right\}|0\rangle
=\delta^{AC}\int\frac{d^4\ell}{(2\pi)^4}\frac{-ig_\perp^{\sigma\rho}}{\ell^2+i0}\;e^{i\ell\cdot y}\;,
\label{eq:Bprop}
\end{equation}
This expression is the first term of the full propagator expression for the collinear gauge-invariant gluon building block $\mathcal{B}_n^\mu$ obtained without any gauge-fixing (see~\cite{Bauer:2008qu} for more details). The other terms in the full expression are proportional to $\bar{n}^\mu$ and thus forced to be zero due to the perp indices of $\mathcal{B}_{n \perp}^\mu$. Furthermore, the derivative on the gluonic building block gives
\begin{equation}
(\bar n\cdot\partial)\,\mathcal B_{n\perp}^{A\sigma}=(i\bar n\cdot\ell)\,\mathcal B_{n\perp}^{A\sigma}\;,
\end{equation}
thereby generating the numerator appearing in Eq.~\eqref{eq:spec1}.

At this stage, all hard and hard-collinear fluctuations have been integrated out perturbatively, leaving behind a matrix element involving only soft and collinear fields. The remaining non-perturbative dynamics is encoded in the light-cone distribution amplitudes of the $B$ and $K$ mesons, allowing the matrix element to be factorized into universal hadronic wave functions convoluted with a perturbatively calculable hard-scattering kernel.

\subsubsection{Factorization of the Spectator Matrix Element}
At leading power in the SCET expansion, the collinear fields in Eq.~\eqref{eq:spec1} are absorbed into the kaon projector, while the soft spectator quark and the heavy-quark field combine to form the $B$-meson projector. To make this factorization explicit, we insert the standard light-cone projectors for the kaon and the $B$ meson. The kaon matrix element is parameterized as
\begin{equation}
\langle K(p_K)|(\bar\xi_{n\alpha} W_n)_i(0)(W_n^\dagger\xi_{n\eta})_l(y)|0\rangle= \frac{\delta_{il}}{N_c}\mathcal{K}_{\alpha\eta}\;,
\end{equation}
where
\begin{equation}
\mathcal{K}_{\alpha\eta}=\frac{i f_K}{4}
(\bar{n}\cdot p_K)\left(\frac{\slashed{n}}{2}\gamma_5\right)_{\eta\alpha}
\int_0^1du\, e^{i\bar u p_K\cdot y} \phi_K(u)\;,
\end{equation}
with $\phi_K(u)$ denoting the leading-twist kaon light-cone distribution amplitude~\cite{Beneke:2003pa}. The color factor $1/N_c$ comes from the singlet color projection of $(\bar{\xi}_n W_n)_i (W_n^\dagger \xi_n)_l$ whereas the color-octet projection gives a vanishing matrix element for meson states. Similarly, the $B$-meson matrix element is written as
\begin{equation}
\langle0|(\bar q_{s\delta}Y_s)_k(y)(Y_s^\dagger h_{v\beta})_j(0) |B(v)\rangle
=
\frac{\delta_{kj}}{N_c}
\mathcal{J}_{\delta \beta}\;,
\end{equation}
where
\begin{equation}
\mathcal{J}_{\delta \beta}
=
-\frac{i f_Bm_B}{4}\left[\frac{1+\slashed{v}}{2}(n.v)\slashed{\bar n}\;\widetilde\phi_{B^+}\gamma_5\right]_{\beta\delta}\;,
\end{equation}
and~\cite{Grozin:1996pq}
\begin{equation}
    \widetilde{\phi}_{B^+}=\int_0^\infty d\omega\,e^{-ik_s\cdot y}\,\phi_B^+(\omega)\;.
\end{equation}
Here $\omega \equiv n.k_s$ is the Fourier momentum corresponding to the spectator $B$-meson quark. Substituting these projectors into Eq.~\eqref{eq:spec1}, separates the spinor and color structures from the non-perturbative hadronic dynamics. The color indices are contracted using the $SU(N_c)$ identity
\begin{equation} 
\frac{\delta_{il}}{N_c} \frac{\delta_{kj}}{N_c} (T^A_{ij}T^A_{kl})=\frac{\delta_{il} \delta_{kj}}{2N_c^2}\left(\delta_{il}\delta_{kj}-
\frac1{N_c}\delta_{ij}\delta_{kl}\right)  \;,
\end{equation}
where both terms contribute. Their sum combines to the familiar factor of $(1-1/N_c^2)/2=C_F/N_c$, leaving a single Dirac trace multiplying the convolution of the two LCDAs.

\subsubsection{Dirac Structure and Momentum-Space Factorization}

Substituting the light-cone projectors for the meson fields into Eq.~\eqref{eq:spec1}, together with the color identity discussed before, the spectator amplitude assumes the form
\begin{align}
&\mathcal{M}_{\rm spec}^{\partial a g}
=
\left(g_sm_B\right)\frac{i\mathcal{C}_{\partial ag}}{2f_a}\frac{C_F}{N_c}\frac{f_Kf_Bm_B}{16}\left(\bar{n}.p_K\right)\left(n.v\right)
\nonumber\\
&
\quad\times\int_0^1du\int_0^\infty d\omega\int d^4y
\int\frac{d^4\ell}{(2\pi)^4}\frac{\bar n\!\cdot\!\ell}{\ell^2+i0}
e^{i(\ell+\bar up_K-k_s)\cdot y}
\nonumber\\
&
\quad\times\phi_K(u)\phi_B^+(\omega)\,T_{\mu\rho}\,\epsilon_\perp^{\mu\rho}\;,
\label{eq:Mfactorized}
\end{align}
where the entire Dirac trace is explicitly given as
\begin{equation}
T_{\mu\rho}={\rm Tr}\left[\frac{\slashed{n}}{2}\gamma_5\gamma_{\perp\mu}P_L\frac{1+\slashed{v}}{2}{\slashed{\bar{n}}}\gamma_5\gamma_{\perp\rho} \right]\;.
\label{eq:Tmunu}
\end{equation}

Using the projectors together with the anticommutation properties of $\gamma_5$, the trace naturally decomposes into parity-even and parity-odd pieces,
\begin{align}
&\qquad \qquad T_{\mu\rho}=\frac18\left(T_{\mu\rho}^{(1)}+T_{\mu\rho}^{(2)}\right)\;, \qquad\text{where}\\
&T_{\mu\rho}^{(1)}={\rm Tr}\left[\gamma_{\perp\mu} \slashed{\bar{n}}\gamma_{\perp\rho}\slashed{n}\right]\;, \quad
T_{\mu\rho}^{(2)}=+{\rm Tr}\left[\gamma_{\perp\mu}\slashed{\bar{n}}\gamma_5\gamma_{\perp\rho}\slashed{n}\right]\;. \nonumber
\end{align}

The evaluation of the Dirac traces is performed using the standard four-dimensional Clifford algebra. The parity-odd contribution, which contains four gamma matrices and $\gamma_5$, yields a Levi-Civita tensor. This considerably simplifies the trace algebra, and one finds
\begin{equation}
    T_{\mu\rho}= -g_{\mu\rho}^{\perp\perp}-i\epsilon_{\rho\mu}^{\perp\perp}\;.
    \label{eq:tensor}
\end{equation}
Notice that Eq.~\eqref{eq:Mfactorized} already has the antisymmetric tensor; therefore, only the antisymmetric part of Eq.~\eqref{eq:tensor} contributes, $\epsilon_{\perp}^{\mu\rho}\epsilon_{\perp\mu\rho}=2$.  The entire spinor structure thus collapses to a scalar coefficient, leaving only the momentum integrations over the hard-collinear propagator and the light-cone distribution amplitudes. 

\subsubsection{Momentum Integrations and Spectator Contribution}

The remaining integrations are purely kinematical and can therefore be carried out analytically. In particular, the integration over the space-time coordinate $y$ enforces overall momentum conservation through the resulting four-dimensional delta function. This fixes the momentum flowing through the hard-collinear propagator to be
\begin{equation}
\ell^\mu=k_s^\mu - \bar u\,p_K^\mu\;,
\end{equation}
where $\bar u\,p_K^\mu$ denotes the momentum carried by the collinear anti-quark inside the kaon and, as mentioned before, $k_s^\mu$ is the momentum of the soft spectator quark in the $B$ meson. Consequently, the internal gluon momentum is completely determined by the external kinematics and the convolution variables appearing in the meson light-cone distribution amplitudes. After substituting this relation into the hard kernel, the spectator-scattering amplitude factorizes into a convolution of perturbatively calculable hard coefficients with the leading-twist light-cone distribution amplitudes of the $B$ and $K$ mesons. The resulting expression can be written as
\begin{align}
&\mathcal M_{\rm spec}^{\partial a g}= \bigg(g_s \frac{\mathcal{C}_{\partial a g}}{f_a} \bigg)\frac{C_F}{N_c}\frac{f_Kf_Bm_B^2 }{16}(\bar n\!\cdot p_K)(n\!\cdot v)
\nonumber\\
&
\times\int_0^1du\int_0^\infty d\omega\,\phi_K(u)\phi_B^+(\omega)\frac{\bar n\!\cdot( k_s - \bar up_K)}{(k_s - \bar up_K)^2+i.0}\;.
\label{eq:Mspecbefore}
\end{align}
The spectator quark carries soft momentum $k_s^\mu=\omega \bar n^\mu/2\;$,
while the energetic kaon momentum is $p_K^\mu=E_Kn^\mu\;,\; \bar n\cdot p_K=2E_K\;.$
Using these relations, we find
\begin{equation}
    \frac{\bar n\!\cdot(k_s - \bar up_K)}{( k_s - \bar up_K)^2+i0}
    =\frac{1}{\omega}\;,
\end{equation}
so that the dependence on the momentum fraction $u$ completely cancels inside the hard-scattering kernel. Furthermore, the leading-twist kaon and $B$-meson LCDA are conventionally defined as~\cite{Lepage:1980fj, Efremov:1979qk, Lange:2003ff}
\begin{equation}
    \int_0^1du\,\phi_K(u)=1\;,\quad  \frac{1}{\lambda_B}\equiv\int_0^\infty\frac{d\omega}{\omega}\phi_B^+(\omega)\;.
\end{equation}

Finally, choosing the rest frame of the $B$ meson and $v^\mu = (n^\mu + \bar{n}^\mu)/2$, Eq.~\eqref{eq:Mspecbefore} yields the following compact form of the spectator contribution
\begin{equation}
    \mathcal{M}_{\rm spec}^{\partial ag} = g_s\, \frac{\mathcal{C}_{\partial a g}}{f_a} \frac{C_F E_K}{N_c} \frac{f_K f_B m_B^2}{8\lambda_B}\;,
    \label{eq:spec}
\end{equation}
up to corrections suppressed by the residual axion momentum.

\subsection{Soft contribution from: \texorpdfstring{$\mathcal{O}_{\partial a g}$}{O[\textbackslash partial a g]}} \label{sec:softOdelag}
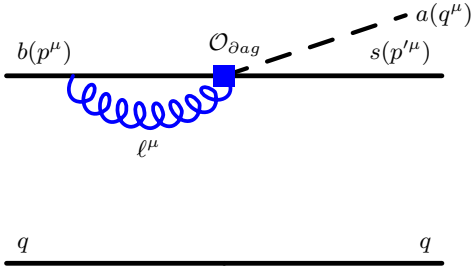
\begin{figure}[h]
	\centering
	\begin{tikzpicture}[line cap=round, line join=round, scale=0.8]
		\coordinate (T)  at (0,0);
		\coordinate (B)  at (0,-1);
		\coordinate (S)  at (0,-3.10);
		\coordinate (Bg) at (-2.5,0);   
		
		\draw[line width=1.8pt] (-3.6,0) node[above right] {\(b(p^\mu)\)} -- (0,0);
		\draw[line width=1.8pt] (0,0) -- (3.6,0) node[above left] {\(s(p'^{\mu})\)};
		\draw[line width=1.8pt] (-3.6,-3.10) node [above right] {$q$} -- (0,-3.10);
		\draw[line width=1.8pt] (0,-3.10)  -- (3.6,-3.10) node [above left] {$q$};
		
		\draw[
		line width=1.6pt,
		dashed,
		dash pattern=on 9pt off 8pt
		] (T) -- (3,1) node[right] {\(a(q^\mu)\)};
		
		\draw[
		line width=1.4pt, blue,
		decorate,
		decoration={coil, aspect=0.65, segment length=7pt, amplitude=5pt}
		] (T) to[out=-120, in=-60] (Bg);
		
		\node at (-1.25,-1.2) {\(\ell^\mu\)};
		
		\fill[blue] (-0.18,-0.18) rectangle (0.18,0.18) node[above, black] {\({\cal O}_{\partial a g}\)};
	\end{tikzpicture}
    \caption{Representative hard scale diagram contributing to the soft overlap matching onto SCET$_{\rm I}$. The gluon (in blue) denotes a hard propagator with virtuality $\sim m_b^2$, whereas the blob indicates the insertion of the $\mathcal{O}_{\partial a g}$ operator. After integrating out the hard gluon, one generates a contact operator with $h_v-\xi_n-a$.}
\label{fig:fig2}
\end{figure}

The soft overlap contribution arises from integrating out hard fluctuations at the $m_b$ scale, thereby matching the low-energy QCD operators onto SCET currents. As mentioned before, the dominant contribution comes from the insertion of $\mathcal{O}_{\partial a g}$ operator, whose representative diagram contributing to $B\to Ka$ is shown in Fig.~\ref{fig:fig2}.
At the matching scale $\mu\sim m_B$, all internal propagators carry hard virtualities of order $m_B^2$. Consequently, the finite part of the one-loop amplitude determines the Wilson coefficient of the operator $\mO_{bsa}^{\rm SCET}$. Throughout this work, the matching is performed in the $\overline{\rm MS}$ scheme, and the ultraviolet poles are subtracted accordingly. Since the matching is carried out directly at the hard scale, no renormalization-group evolution between two different scales is considered at this stage.
 
Similar to Fig.~\ref{fig:fig2}, another possible diagram is where the hard gluon from the hard vertex attaches back on the $s$-quark. The denominators corresponding to that case are $\ell^2$ and $(p'-\ell)^2-m_s^2$. For $m_s\to0$ and $p'^2=0$ the integral becomes scaleless and vanishes in dimensional regularization with $1/\epsilon_{\rm UV}-1/\epsilon_{\rm IR}=0$. The corresponding diagram in SCET$_\text{I}$ is also scaleless for the same kinematic reason. As a result, this diagram does not contribute in matching to $\mO_{bsa}^{\rm SCET}$. Therefore, at the leading order, the only relevant diagram contributing to the soft form factor is the one shown in Fig.~\ref{fig:fig2}.

\subsubsection{Radiatively Generated Operator and Matching to SCET}
The quark-level loop amplitude can be expressed as
\begin{align}
    &\mathcal{A}_{\partial ag} = \frac{\mC_{\partial a g}}{f_a} \left(i g_s C_F \epsilon_{\mu\nu\rho\sigma}\right)\nonumber \\
    &\int \frac{d^d\ell}{(2\pi)^d} \frac{\bar{u}(p^\prime)\gamma^\mu P_L\left(\slashed{p}-\slashed{\ell}+m_b\right)\gamma^\sigma u(p)q^\nu\ell^\rho}{\ell^2 \left[(p-\ell)^2-m_b^2\right]}\;.
\end{align}
The steps are rather straightforward, and it generates
\begin{align}
\mathcal{A}_{\partial a g} =&i\left(\frac{\mathcal{C}_{\partial a g} g_s C_F}{16\pi^2 f_a}\right)\frac{(17 m_b^2- 2m_a^2)}{9}\;\bar{u}(p^\prime)\slashed{q}P_L u(p)\;.
\end{align}

The final step in the matching from QCD to SCET is to express the resulting spinor structure $\bar{u}(p^\prime)\slashed{q}P_L u(p)$ onto the corresponding SCET heavy-to-light currents with two-component spinors. At the leading power~\cite{Bauer:2000yr}, 
\begin{equation}
\begin{aligned}
     q_\mu \bar{u}(p') \gamma^\mu P_L u(p) &\to \frac{1}{2} \bar{\chi}_n \slashed{q}_\perp h_v + (n.q) \bar{\chi}_n P_R h_v \\
     &- \frac{1}{2} iq_\mu \epsilon_\perp^{\mu\nu} \bar{\chi}_n \gamma_\nu^\perp h_v\;.
\end{aligned}
\end{equation}
Therefore, we find that $\mathcal{O}_{\partial ag}$ matches onto the following $\text{SCET}_{\text{I}}$ operator
\begin{equation}
\mathcal{O}_{bsa}^{\text{SCET}} = \frac{\partial_\mu a}{2} \bar{\chi}_n \left( \gamma^\mu_\perp + n^\mu (1+\gamma_5) - i\epsilon^{\mu\nu}_\perp \gamma_\nu^\perp\right) h_v\;.
    \label{eq:dirac_str}
\end{equation}
with the Wilson coefficient evaluated from the loop diagram shown in Fig.~\ref{fig:fig2},
\begin{align}
\frac{1}{f_a} \mathcal{C}^{\rm SCET}_{bsa - \partial a g} =\frac{\mathcal{C}_{\partial a g}\,g_s\,C_F\,}{16\pi^2 f_a}
\left[\frac{17m_b^2- 2m_a^2}{9}\right]\;.
\end{align}

There are four independent Dirac structures above: $\{1, \gamma_5, \gamma_\mu^\perp \}$ (the perp index $\mu$ takes on two values). However, for $B\to P$ transitions only the pure scalar structure, i.e., the identity matrix in spinor space, gives a non-vanishing contribution~\cite{Bauer:2000yr, Chay:2002vy}. This nonperturbative matrix element is parameterized as the universal soft form factor in SCET~\cite{Charles:1998dr, Beneke:2000wa}
\begin{equation}
    \langle K(p')| \bar{\chi}_n h_v|B(p)\rangle = 2E_K \zeta(E_K)\;.
\end{equation}
We thus obtain the soft-overlap contribution of $\mathcal{O}_{\partial a g}$ to the meson-level $B\to Ka$ amplitude as
\begin{equation}
    \mathcal{M}^{\partial a g}_{\text{soft}} = \frac{\mathcal{C}_{bsa - \partial a g}^{\text{SCET}}}{f_a} (n.q )\; E_K\; \zeta(E_K)\;.
    \label{eq:soft}
\end{equation}
The matrix element is expressed directly in terms of the soft-overlap form factor $\zeta(E_K)$, consistent with the discussion around Eq.~\eqref{eq:f0split}.

In the following section, we discuss the contribution to $B\to Ka$ from the chromomagnetic operator $\mO_{8g}$. However, we note that the numerically dominant contributions come from the diagrams induced by $\mO_{\partial ag}$. This follows from the hierarchy of the Wilson coefficients established in the previous section, where $\mC_{Dg}$ is found to be substantially larger than $\mC_{8g}$. Furthermore, among the charged-current four-quark operators $\mO_{1-2}$, the dominant effect is from the charm loop, which is already accounted for in $E_0$ as discussed in Sec.~\ref{sec:sec2}. The remaining contributions from $\mO_{3-6}$ arise only through additional loop-induced hard matching and are therefore further suppressed by a factor of $\alpha_s/(4\pi)$. As a result, the matching generated by $O_{\partial ag}$ provides the dominant contribution to the soft-overlap amplitude, while the remaining operators constitute subdominant corrections.

\section{Contribution from the Chromomagnetic Operator: \texorpdfstring{$\mathcal{O}_{8g}$}{O8g}}
\label{sec:sec4}

\subsection{Spectator contribution from: $\mathcal{O}_{8g}$}

	\newcommand{\Diagmatching}{
	\begin{tikzpicture}[line cap=round, line join=round, scale=0.8]
		\coordinate (T)  at (0,0);
		\coordinate (B)  at (0,-1);
		\coordinate (mid)  at (0,-1.5);
		\coordinate (S)  at (0,-3.10);
		\coordinate (Bg) at (-2.5,0);   
		
		\draw[line width=1.8pt] (-3.6,0) node[above right] {\(b(p^\mu)\)} -- (0,0);
		\draw[line width=1.8pt] (0,0) -- (3.6,0) node[above left] {\(s(p'^\mu)\)};
		\draw[line width=1.8pt] (-3.6,-3.10) node[above right] {\(q\)} -- (0,-3.10);
		\draw[line width=1.8pt] (0,-3.10) -- (3.6,-3.10) node[above left] {\(q\)} ;
		
		\draw[
		line width=1.6pt,
		dashed,
		dash pattern=on 9pt off 8pt
		] (mid) -- (2.7,-1.5) node[right] {\(a(q^\mu)\)};
		
		\draw[
		line width=1.4pt, blue,
		decorate,
		decoration={coil, aspect=0.65, segment length=7pt, amplitude=5pt}
		] (T) to (mid) node[left, black] {\({\cal O}_{agg}\)};
		\draw[
		line width=1.4pt, purple,
		decorate,
		decoration={coil, aspect=0.65, segment length=7pt, amplitude=5pt}
		] (mid) to (S) ;
		
        \node at (.7,-.8) {\(k^\mu\)};
        \node at (.7,-2.2) {\(\ell^\mu\)};
		
		\fill[blue] (-0.18,-0.18) rectangle (0.18,0.18)
		node[above, black] {\({\cal O}_{8g}\)} ;
		\fill[black] (-0.1,-1.6) rectangle (.1,-1.4);
	\end{tikzpicture}
}

\newcommand{\DiagDipoleScet}{
	\begin{tikzpicture}[line cap=round, line join=round, scale=0.8]
		\coordinate (T)  at (0,0);
		\coordinate (B)  at (0,-1);
		\coordinate (mid)  at (0,-1.5);
		\coordinate (S)  at (0,-3.10);
		\coordinate (Bg) at (-2.5,0); 
		\draw[line width=1.8pt] (-3.6,0) node[above right] {\(h_v (p_b^\mu)\)} -- (0,0);
		\draw[line width=1.8pt] (0,0) -- (3.6,0) node[above left] {\(\xi_n(u p_K^\mu)\)};
		\draw[line width=1.8pt] (-3.6,-3.10) node[above right] {\(q_s(k_s^\mu)\)} -- (0,-3.10);
		\draw[line width=1.8pt] (0,-3.10) -- (3.6,-3.10) node[above left] {$q_n(\bar u p_K^\mu)$};
		
		\draw[
		line width=1.6pt,
		dashed,
		dash pattern=on 9pt off 8pt
		] (T) -- (2.7,1) node[right] {\(a(q^\mu)\)};
		
		\draw[
		line width=1.4pt, purple,
		decorate,
		decoration={coil, aspect=0.65, segment length=7pt, amplitude=5pt}
		] (T) to (S);
		\node at (.8,-1.4) {\(\ell^\mu\)};
		
		\fill[gray] (-0.18,-0.18) rectangle (0.18,0.18)
		node[above, black] {\({\cal O}_{8ag}\)} ;
	\end{tikzpicture}
}
\begin{figure}[h]
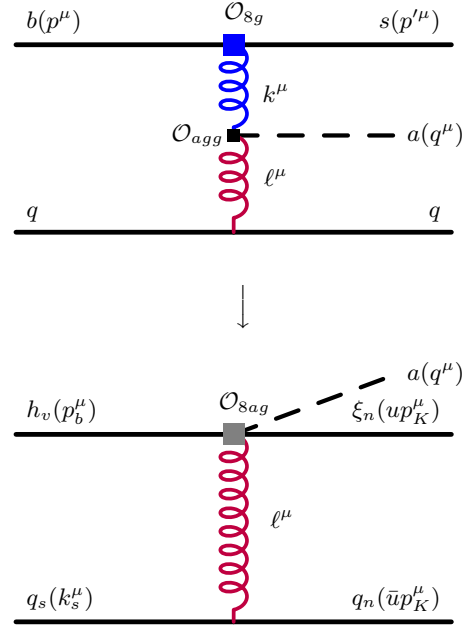

	\centering
 \Diagmatching \\
     $$
     \Big\downarrow
     $$
 \DiagDipoleScet
 \caption{$\mathcal{O}_{8g}$ and $aG\tilde{G}$ derived contribution to spectator scattering. In the top diagram, the blue gluon denotes a hard propagator with virtuality $\sim m_b^2$, whereas the blob indicates $\mathcal{O}_{8 g}$ insertion. The top diagram matches onto the bottom diagram after integrating out the hard gluon and gives the spectator contribution from the effective $\mathcal{O}_{8ag}$ operator.}
\label{fig:fig3}
\end{figure}

The spectator contribution descending from the operators $\mathcal{O}_{8g}$ with $aG\tilde{G}$ and shown in Fig.~\ref{fig:fig3}. This contribution is predicted to be small compared to $\mathcal{O}_{\partial a g}$ because of the smaller Wilson coefficient of the governing operator. The WET operators $Q_3 - Q_6$'s contribution is further loop-suppressed compared to $\mathcal{O}_{8g}$ and will not be considered. The external momenta in the above tree-level diagram follow the same scalings as before, shown in Table~\ref{tab:scaling8ga}. This fixes the scaling of the lower gluon (shown in red) as $n$-hard-collinear. However, the scaling of the upper gluon turns out to be $k\sim (1,1,\lambda^2)$. Being off-shell with hard virtuality $\sim m_b^2$, this gluon is integrated out in the matching between QCD and SCET at $\mu\sim m_b$.
The tree-level matching of $\mathcal{O}_{8g}$ and $aG\tilde{G}$ insertions with the hard internal gluon integrated out gives the following $\text{SCET}_\text{I}$ operator at leading power 
\begin{equation}
    \mathcal{O}_{8ag}^{\text{SCET}} = m_b a(x)[ \bar{\chi}_n \gamma^\mu_\perp P_L T^A Y_s^\dagger h_v] \epsilon^\perp_{\mu\nu} (\bar{n}.\partial )\mathcal{B}^{A \nu}_{n\perp}\;,\label{eq:O8g}
\end{equation}
with the Wilson coefficient
\begin{equation}
   \frac{\mathcal{C}_{8ag}^{\text{SCET}}}{f_a} = \frac{4\mathcal{C}_{gg}}{f_a} \mathcal{C}_{8g}  \bigg(\frac{1}{1-\bar{n}.\hat{p}} \bigg)\;,
\end{equation}
where $\hat{p} \equiv p/m_b$ and $p^\mu$ is the light quark momentum.

The spectator scattering contribution of $\mathcal{O}_{8ag}^{\text{SCET}}$ (bottom panel in Fig.~\ref{fig:fig3}) is found to be

\begin{equation}\label{eq:8agspec}
     \mathcal{M}_{\rm spec}^{8ag} = \frac{\mathcal{C}_{gg} \mathcal{C}_{8g}g_s}{f_a}\frac{C_F E_K}{N_c} \frac{3f_K f_B m_B m_b}{\lambda_B}\;.
\end{equation}

Naively, another type of spectator contribution seems possible from the time-ordered products of two SCET operators: 
\begin{enumerate}
    \item $\mathcal{O}_{8g}^{\text{SCET}}$ with the gluon being $\bar{n}$ hard-collinear~\cite{Hurth:2023paz}.
    \item $\mathcal{O}_{agg}^{\text{SCET}}$ with an $\bar{n}$ and a $n$-hard-collinear gluon.
\end{enumerate}
Both operators will appear upon matching to SCET at the hard scale $m_b$. The $\bar{n}$ hard-collinear gluons will contract, while the $n$ hard-collinear gluon will attach to the spectator quark line as in the previous cases. However, the scaling rules as shown in Table~\ref{tab:scaling8ga} imply that the upper internal gluon has \textit{hard} momenta scaling from external kinematics, rather than $\bar{n}$ hard-collinear. For the latter, one should have $\bar{n}.k = \bar{n}.p_b - u\bar{n}.p_K = m_b - u2E_K \sim \lambda^2$ but this involves a fine-tuned cancellation between two $\lambda^0$ numbers and spoils manifest power counting. Taking $u, \bar{u} \sim 1$, $\bar{n}.k = m_b(1-x) \sim \lambda^2$ is only possible if $x \to 1$, i.e. the axion is $\bar{n}$-collinear: $q^\mu \sim (\lambda^4,1, \lambda^2)$ (see Sec.~\ref{subsubsec:spectatorcontribution}). Considering this leads to a contradiction since now the lower internal gluon's scaling becomes inconsistent: $l^\mu = q^\mu - k^\mu \Rightarrow l^2 \sim (q-k)^2$, but $(q-k)^2 \sim \lambda^0$. The lower internal gluon cannot be hard, since it connects to the light-quark spectator line. Hence we conclude that such a spectator contribution is forbidden by kinematics.

\begin{table}[h]
\centering
\renewcommand{\arraystretch}{1.4}
\begin{tabular}{|l|l|l|}
\hline
Momentum & Scaling & Mode \\
\hline
\hline
$p_b^\mu=m_bv^\mu+r^\mu$ & $(1,1,\lambda^2)$ & Heavy \\
$k_s^\mu$ & $(\lambda^2,\lambda^2,\lambda^2)$ & Soft \\
$p_K^\mu$ & $(\lambda^4,1,\lambda^2)$ & $n$-collinear \\
$\ell^\mu=k_s^\mu-\bar u\,p_K^\mu$ & $(\lambda^2,1,\lambda^2)$ & $n$-hard collinear \\
$k^\mu=m_Bv^\mu+r^\mu-u\,p_K^\mu$ & $(1,1,\lambda^2)$ & Heavy/NR \\
$q^\mu=k^\mu+\ell^\mu$ & $(1,1,\lambda^2)$ & Heavy/NR \\
\hline
\end{tabular}
\caption{Momentum assignments and corresponding SCET power counting for the process in Fig.~\ref{fig:fig3}. }
\label{tab:scaling8ga}
\end{table}

\subsection{Soft contribution from: $\mathcal{O}_{8g}$}
\begin{figure}[h]
	\centering
\begin{tikzpicture}[line cap=round, line join=round, scale=0.8]
		\coordinate (T)  at (0,0);
		\coordinate (B)  at (0,-1);
		\coordinate (S)  at (0,-3.10);
		\coordinate (Bg) at (-2.5,0);   
		\coordinate (Mid) at (-1.25,-0.95); 
		
		\draw[line width=1.8pt] (-3.6,0) node[above right] {\(b(p^\mu)\)} -- (0,0);
		\draw[line width=1.8pt] (0,0) -- (3.6,0) node[above left] {\(s(p'^{\mu})\)};
		\draw[line width=1.8pt] (-3.6,-3.10) node [above right] {$q$} -- (0,-3.10);
		\draw[line width=1.8pt] (0,-3.10)  -- (3.6,-3.10) node [above left] {$q$};

        \node at (-1.2, 0.4) {\(k^\mu\)};
        
		\draw[
		line width=1.4pt, blue,
		decorate,
		decoration={coil, aspect=0.65, segment length=7pt, amplitude=5pt}
		] (T) to[out=-120, in=-60] (Bg);
		
		\draw[
		line width=1.6pt,
		dashed,
		dash pattern=on 9pt off 8pt
		] (Mid) -- ++(1.4,-1.2) node[right] {\(a(q^\mu)\)};

		\fill[blue] (-0.18,-0.18) rectangle (0.18,0.18) node[above, black] {\({\cal O}_{8g}\)};
	\end{tikzpicture}
    \caption{Representative hard scale diagram contributing to the soft overlap matching onto SCET$_{\rm I}$. The blob indicates the insertion of the $\mathcal{O}_{8 g}$ operator. After integrating out the hard gluon, one generates a contact operator with $h_v-\xi_n-a$.}
\label{fig:8gsoft}
\end{figure}
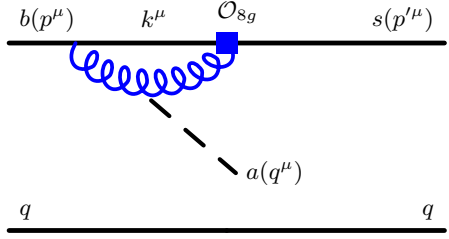

The chromomagnetic operator also gives a soft-overlap contribution to $B\to Ka$ when it is combined with $aG\widetilde{G}$ as illustrated in the diagram of Fig.~\ref{fig:8gsoft}. The other possible diagram with the gluon attaching to the light quark, and the diagrams in SCET$_\text{I}$, are excluded by the same arguments discussed in Section~\ref{sec:softOdelag}. The total quark-level amplitude of Fig.~\ref{fig:8gsoft} is given by
\begin{equation}
\begin{aligned}
    & \mathcal{A}_{8g} = \frac{\mathcal{C}_{gg}}{f_a}(8 g_s \mathcal{C}_{8g} m_b)C_F\ \epsilon_{\eta\xi\rho\sigma}q^\sigma g^{\nu \xi} 
      \int \frac{d^d k}{(2\pi)^d} \\
      & \frac{ \bar{u}(p') \sigma_{\mu\nu}P_R (p-q-k)^\mu (\slashed{k} + m_b) (p-k)^\rho \gamma^{\eta} u(p)}{ [k^2 - m_b^2][ p-k-q ]^2 [p-k]^2 }\;.
\end{aligned}
\end{equation}
The various spinor structures generated by the Dirac algebra can all be reduced to the left-handed $b\to sa$ operator structure, as in the case of $\mathcal{O}_{\partial a g}$, by using momentum conservation 
and the spinor equation of motion for the $b$ and $s$ quarks. We have also checked that the Dirac structure in Fig.~\ref{fig:8gsoft} does not generate any evanescent operators~\cite{Bisht:2026qwe}. As in Section~\ref{sec:softOdelag}, the operator generated by matching the spinor structure to SCET at leading power is the one given in Eq.~\eqref{eq:dirac_str} with the Wilson coefficient

\begin{equation}
\begin{aligned}
    & \frac{1}{f_a} \mathcal{C}_{bsa-8g}^{\text{SCET}} = \bigg(\frac{\mathcal{C}_{8g}g_s \mathcal{C}_{gg} C_F  }{ 4\pi^2 f_a} \bigg)\\
     &\times \bigg[ 3(m_b^2 + m_a^2) +i\pi m_a^2  + m_a^2 \log\bigg(\frac{m_b^2}{m_a^2} \bigg) \bigg] \;.
\end{aligned}
\end{equation}
The corresponding $B\to Ka$ soft-overlap amplitude is
\begin{equation}
     \mathcal{M}^{8g}_{\text{soft}} = \frac{\mathcal{C}_{bsa-8g}^{\text{SCET}}}{f_a} (n.q )\; E_K\; \zeta(E_K)\;.
    \label{eq:soft8g}
\end{equation}

\section{Implication of the direct operator: \texorpdfstring{$\mathcal{O}_{bsa}$}{Obsa}}\label{sec:sec5}
\begin{figure}[h]
\centering
\resizebox{3.85cm}{!}{%
\begin{tikzpicture}
    \begin{feynman}
    \vertex (a) {\(b\)};
    \vertex [right=1cm of a] (b);
    \vertex [right=0.6cm of b] (c);
    \vertex [right=1cm of c] (d);
    \vertex [right=0.6cm of d] (e);
    \vertex [right=1cm of e] (f) {\(s\)};
    \vertex at ($(c)!0.5!(d)!1.08!135:(c)$) (g);
    \vertex [below right=0.8cm of g] (h) {\(a\)};
    \vertex [below = 1.55cm of a] {};
    \diagram* {
            (a) -- [fermion] (b) -- [fermion] (c) -- [fermion] (d) -- [fermion] (e) -- [fermion] (f), (b) -- [boson, half left, color=blue, style={line width=1pt}] (e), (c) -- [gluon, half right, color=blue, style={line width=1pt}] (d), (g) -- [scalar] (h)
        };
        \end{feynman}
\end{tikzpicture}
}
\resizebox{3.85cm}{!}{%
\begin{tikzpicture}
    \begin{feynman}
    \vertex (a) {\(b\)};
    \vertex [right=1cm of a] (b);
    \vertex [right=0.6cm of b] (c);
    \vertex [right=1cm of c] (d);
    \vertex [right=0.6cm of d] (e);
    \vertex [right=1cm of e] (f) {\(s\)};
    \vertex at ($(b)!0.5!(e)!0.9!135:(b)$) (g);
    \vertex [below right=0.8cm of g] (h) {\(a\)};
    \diagram* {
            (a) -- [fermion] (b) -- [fermion] (c) -- [fermion] (d) -- [fermion] (e) -- [fermion] (f), (c) -- [boson, half left, color=blue, style={line width=1pt}] (d), (b) -- [gluon, half right, looseness = 1.2, color=blue, style={line width=1pt}] (e), (g) -- [scalar] (h)
        };
        \end{feynman}
\end{tikzpicture}
}
\\
\resizebox{3.85cm}{!}{%
\begin{tikzpicture}
    \begin{feynman}
    \vertex (a) {\(b\)};
    \vertex [right=1cm of a] (b);
    \vertex [right=0.7cm of b] (c);
    \vertex [right=0.8cm of c] (d);
    \vertex [right=0.7cm of d] (e);
    \vertex [right=1cm of e] (f) {\(s\)};
    \vertex at ($(c)!0.5!(e)!1.05!135:(c)$) (g);
    \vertex [below right=0.8cm of g] (h) {\(a\)};
    \diagram* {
            (a) -- [fermion] (b) -- [fermion] (c) -- [fermion] (d) -- [fermion] (e) -- [fermion] (f), (b) -- [boson, half left, color=blue, style={line width=1pt}] (d), (c) -- [gluon, half right, color=blue, style={line width=1pt}] (e), (g) -- [scalar] (h)
        };
        \end{feynman}
\end{tikzpicture}
}
\resizebox{3.85cm}{!}{%
\begin{tikzpicture}
    \begin{feynman}
    \vertex (a) {\(b\)};
    \vertex [right=1cm of a] (b);
    \vertex [right=0.7cm of b] (c);
    \vertex [right=0.8cm of c] (d);
    \vertex [right=0.7cm of d] (e);
    \vertex [right=1cm of e] (f) {\(s\)};
    \vertex at ($(b)!0.5!(d)!1.05!135:(b)$) (g);
    \vertex [below right=0.8cm of g] (h) {\(a\)};
    \diagram* {
            (a) -- [fermion] (b) -- [fermion] (c) -- [fermion] (d) -- [fermion] (e) -- [fermion] (f), (c) -- [boson, half left, color=blue, style={line width=1pt}] (e), (b) -- [gluon, half right, color=blue, style={line width=1pt}] (d), (g) -- [scalar] (h)
        };
        \end{feynman}
\end{tikzpicture}
}
\caption{Two-loop diagrams generating $\mathcal{O}_{bsa}$ at the electroweak scale by gluonic coupling of the axion. The blue color represents the hard virtuality ($\mu \sim M_W$) that is integrated out to generate the low-energy effective operator $\mathcal{O}_{bsa}$.}
\label{fig:twoloops}
\end{figure}
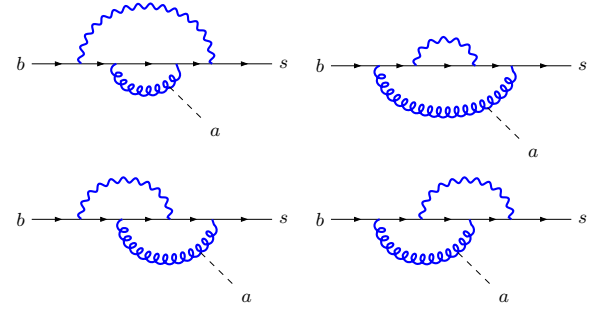

In this section, we consider the more realistic scenario in which the axion-gluon operator $aG\widetilde{G}$ is generated above the electroweak scale. In a consistent effective field theory, however, $aG\widetilde{G}$ cannot be treated in isolation. Renormalizability requires the inclusion of the complete operator basis at a given order, and even if the Wilson coefficients of the remaining operators vanish at the ultraviolet matching scale, renormalization-group evolution inevitably regenerates them through operator mixing. Consequently, electroweak evolution and matching induce additional flavor-changing axion operators at low energies. This realization is therefore of greater phenomenological relevance, as it gives rise to direct $b\to sa$ interactions alongside the gluonic coupling (see Fig.~\ref{fig:twoloops} and Ref.~\cite{ Chakraborty:2021wda, Bauer:2021mvw, Bisht:2024hbs} for details)
\begin{equation}
    \mathcal{O}_{bsa} = \frac{\partial_\mu a}{f_a} \bar{s}\gamma^\mu P_L b\;,
\end{equation}
with a two-loop Wilson coefficient $\mathcal{C}_{bsa}(M_W)$, consistent with the heavy QCD axion scenario is given by 
\begin{equation}
     \mathcal{C}_{bsa} = \frac{\alpha_s}{4\pi} \frac{\alpha_w}{16\pi} \mathcal{C}'^{sw}_{4L}\;,
\end{equation}
where, the explicit expression for $C'^{sw}_{4L}$ is~\cite{Bisht:2024hbs}
\begin{widetext}
\begin{equation}
    \begin{aligned}
        \mathcal C'^{sw}_{4L}(\mu)  &= 6 C_F \sum_{i} V^\ast_{is}V_{ib}\; \mathcal{C}_{gg}(\mu)\Bigg[\log \bigg(\frac{\mu^2}{M_W^2} \bigg) \left(\frac{2 \xi_i  ((\xi_i -2) \xi_i +4) \log (\xi_i )}{(\xi_i -1)^2}-\frac{(\xi_i +2) (3 \xi_i -1)}{\xi_i -1}\right)\\
    & +\frac{\left(6 (3 \xi_i +2) \xi_i ^3-24 \xi_i ^2+12 (\xi_i -2) (\xi_i  (3 \xi_i -1)+1) \log (\xi_i -1)\right)
   \log (\xi_i )}{6 (\xi_i -1)^2 \xi_i }\\
   &-\frac{2 (\xi_i -2) (\xi_i  (3 \xi_i -1)+1) \text{Li}_2\left(\frac{1}{\xi_i
   }\right)}{(\xi_i -1)^2 \xi_i }-\frac{(\xi_i  (\xi_i  (\xi_i +5)-2)+4) \log ^2(\xi_i )}{(\xi_i -1) \xi_i } \\
   & +\frac{\pi ^2 ((\xi_i -2) \xi_i  (2 \xi_i -1)-2)}{3 (\xi_i -1)^2 \xi_i } +\frac{15 \xi_i  (\xi_i  (2 \xi_i -1)+1)-12 (\xi_i +2) (\xi_i  (2 \xi_i -1)+1) \text{Li}_2\left(\frac{\xi_i -1}{\xi_i }\right)}{6 (\xi_i -1) \xi_i }\Bigg]\;.
    \end{aligned}
\end{equation}
\end{widetext}
Here $\xi_i \equiv m_i^2/M_W^2$ and $i = \{u,c,t \}$ are the up-type quarks. In addition, $\mathcal{C}_{bsa}$ also includes dependencies from one-loop generated $\mathcal{C}_{aqq}$ and tree-level $\mathcal{C}_{bsa}$ coefficients, if present at the UV.

\begin{figure}[h]
\centering
\begin{tikzpicture}[line cap=round, line join=round, scale=0.8]
		\coordinate (T)  at (0,0);
		\coordinate (B)  at (0,-1);
		\coordinate (mid)  at (0,-1.5);
		\coordinate (S)  at (0,-3.10);
		\coordinate (Bg) at (-2.5,0); 
		\draw[line width=1.8pt] (-3.6,0) node[above right] {\(h_v (p_b^\mu)\)} -- (0,0);
		\draw[line width=1.8pt] (0,0) -- (3.6,0) node[above left] {\(\xi_n(u p_K^\mu)\)};
		\draw[line width=1.8pt] (-3.6,-3.10) node[above right] {\(q_s(k_s^\mu)\)} -- (0,-3.10);
		\draw[line width=1.8pt] (0,-3.10) -- (3.6,-3.10) node[above left] {$q_n(\bar u p_K^\mu)$};
		
		\draw[
		line width=1.6pt,
		dashed,
		dash pattern=on 9pt off 8pt
		] (T) -- (2.7,1) node[right] {\(a(q^\mu)\)};
		
		\draw[
		line width=1.4pt, purple,
		decorate,
		decoration={coil, aspect=0.65, segment length=7pt, amplitude=5pt}
		] (T) to (S);
		\node at (.8,-1.4) {\(\ell^\mu\)};
		
		\fill[gray] (-0.18,-0.18) rectangle (0.18,0.18)
		node[above, black] {\({\cal O}_{bsa}\)} ;
	\end{tikzpicture}
        \vskip 0.6cm
\begin{tikzpicture}[line cap=round, line join=round, scale=0.8]
		\coordinate (T)  at (0,0);
		\coordinate (B)  at (0,-1);
		\coordinate (mid)  at (0,-1.5);
		\coordinate (S)  at (0,-3.10);
		\coordinate (Bg) at (-2.5,0); 
		\draw[line width=1.8pt] (-3.6,0) node[above right] {$h_v(p_b^\mu)$} -- (0,0);
		\draw[line width=1.8pt] (0,0) -- (3.6,0) node[above left] {\(\xi_n(u p_K^\mu)\)};
		\draw[line width=1.8pt] (-3.6,-3.10) node[above right] {\(q_s (k_s^\mu)\)} -- (0,-3.10);
		\draw[line width=1.8pt] (0,-3.10) -- (3.6,-3.10) node[above left] {\(q_n(\bar u p_K^\mu)\)};
		
		\draw[
		line width=1.6pt,
		dashed,
		dash pattern=on 9pt off 8pt
		] (T) -- (2.7,1) node[right] {\(a(q^\mu)\)};
		
		\fill[gray] (-0.18,-0.18) rectangle (0.18,0.18)
		node[above, black] {\({\cal O}_{bsa}\)} ;
	\end{tikzpicture}
\caption{Leading SCET contribution to $B\to Ka$ derived from the operator $\mathcal{O}_{bsa}$. The top (spectator) and bottom (soft) diagrams are mediated by SCET operators $\mathcal{O}_{bsa}^{(A)\text{SCET}}$ and $\mathcal{O}_{bsa}^{(B)\text{SCET}}$, respectively.}
\label{fig:Obsaspec}
\end{figure}
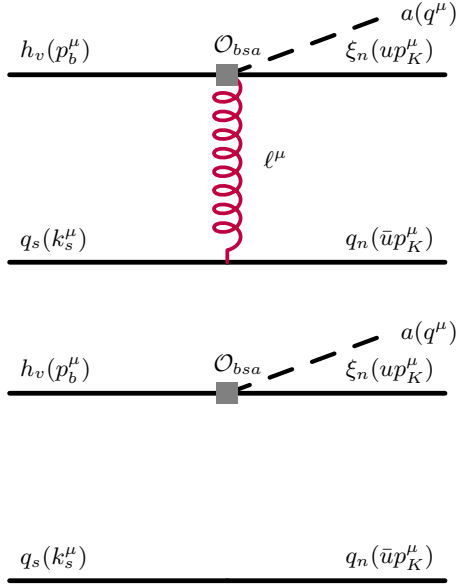

Below the electroweak scale, this operator does not mix with other ALP operators and consequently the Wilson coefficient $\mathcal{C}_{bsa}$  does not run from $M_W$ to $m_B$.  To obtain the spectator and soft effects below the $m_B$ scale descending from this operator (see Fig.~\ref{fig:Obsaspec}), we perform its matching onto SCET. The QCD heavy-to-light current $\bar{s} \gamma^\mu P_L b$ in this operator matches to $A$ and $B$-type currents in SCET~\cite{Beneke:2002ph, Hill:2004if}. The former contains the heavy and light quarks; the latter contains, in addition, a hard-collinear gluon and is subleading in SCET power-counting. However, both $A$ and $B$-type currents contribute at the same $\lambda$-power counting in heavy-to-light decay at the matrix element level~\cite{Bauer:2002aj,Beneke:2003pa}. The aim of this section is to calculate these effects from the QCD operator $\mathcal{O}_{bsa}$.

\subsection{Spectator contribution}
Spectator contributions arise from SCET heavy-to-light currents with a hard-collinear gluon building block $\mathcal{B}$. These $B$-type currents are subleading compared to the $A$-type current because $\mathcal{B}_\perp$ is power-suppressed in $\lambda$. Using the tree-level matching coefficients of QCD currents onto the $B$-type currents from the literature~\cite{Beneke:2004rc, Beneke_2006}, we obtain
\begin{equation}
    \begin{aligned}
        \mathcal{O}^{(B)\text{SCET}}_{bsa} &= -\bigg(\frac{1}{m_b} \bigg) \frac{n.\partial a}{f_a} (\bar{\xi}_n W_n) P_R\slashed{\mathcal{B}}_{n\perp} h_v \\
        &- \frac{1}{2E_K} \frac{\bar{n}.\partial a}{f_a}(\bar{\xi}_n W_n) P_R \slashed{\mathcal{B}}_{n\perp} h_v \\
        & +  \frac{1}{2E_K} \frac{\partial_\mu a}{f_a} (\bar{\xi}_n W_n) P_R \slashed{\mathcal{B}}_{n\perp} \gamma^\mu_\perp h_v\;.\label{eq:Obsa_B}
    \end{aligned}
\end{equation}
Since the NR axion momentum scales as $q^\mu \sim (1,1,\lambda^2)$, the last operator term is subleading and ignored in the following. From the time-ordered product of $\mathcal{O}^{(B)\text{SCET}}_{bsa}$ and $\mathcal{L}_{\xi q G}$ inserted between $B, K$  and $a$ states, we obtain the matrix element
\begin{equation}
 \mathcal{M}^{bsa}_{\text{spec}} =   \frac{\mathcal{C}_{bsa}}{f_a} \frac{C_F}{N_c} (\pi \alpha_s)\frac{3f_K f_B m_B}{2\lambda_B} \bigg( \frac{n.q}{m_b} + \frac{\bar{n}.q}{2E_K} \bigg)\;.
\end{equation}
We simplify this by taking $m_b \simeq m_B$, $n.q \simeq m_B$ and $\bar{n}.q = m_B - 2E_K$ up to power corrections in $\lambda$. This gives
\begin{equation}
    \mathcal{M}^{bsa}_{\text{spec}} =   \frac{\mathcal{C}_{bsa}}{f_a} \frac{C_F}{N_c} (\pi \alpha_s)\frac{3f_K f_B m_B}{2\lambda_B}\frac{m_B}{2E_K} \;.
 \label{eq:spec2}
\end{equation}
We have verified the above expression by substituting the form factor expressions calculated in~\cite{Beneke_2006} into the appropriate QCD vector current matrix element.

\subsection{Soft contribution}
The one-loop matching discussed in subsection~\ref{sec:softOdelag} generates the SCET$_{\rm I}$ operator $\mathcal{O}_{bsa}^{\rm SCET}$, whose matrix element gives the leading soft-overlap contribution to the decay amplitude. Since the hard fluctuations have already been integrated out, the remaining hadronic dynamics are entirely described by matrix elements of the leading-power $A$-type SCET currents. The matching of the operator $\mathcal{O}_{bsa}$ onto this basis is straightforward~\cite{Bauer:2000yr,Beneke:2004rc},
\begin{align}
\mathcal{O}^{(A)\text{SCET}}_{bsa} &= \frac{n.\partial a}{2f_a} (\bar{\xi}_n W_n) (1+\gamma_5)h_v + \frac{\partial_\mu a}{2f_a} (\bar{\xi}_n W_n) \gamma^\mu_\perp h_v \nonumber \\
     & - i\epsilon_\perp^{\mu\nu} \frac{\partial_\mu a }{2f_a} (\bar{\xi}_n W_n) \gamma^\perp_\nu h_v\;. 
     \label{eq:Obsa_A}
\end{align}
The soft-overlap contribution is obtained by taking the hadronic matrix element of $\mathcal{O}^{(A)\rm SCET}_{bsa}$. As discussed below Eq.~\eqref{eq:soft}, only the scalar operator proportional to $(n\!\cdot\!\partial a)(\bar{\xi}_nW_n)h_v$ contributes to the pseudoscalar $B\to K$ transition at leading power, while the transverse Dirac structures vanish. The remaining matrix element is identical to the conventional soft-overlap contribution and is therefore expressed in terms of the soft form factor $\zeta$, as in Eq.~\eqref{eq:soft}. We find
\begin{equation}
\mathcal{M}^{bsa}_{\text{soft}} = \frac{\mathcal{C}_{bsa}}{f_a}\,(n.q)\; E_K\,\zeta(E_K)\big|_{q^2=m_a^2}\;,
\label{eq:soft1}
\end{equation}
which is also in agreement with the literature~\cite{Beneke:2000wa, Beneke_2006}.
This soft contribution has the same form as the one obtained from the $\mathcal{O}_{\partial a g}$ operator in Eq.~\eqref{eq:soft}. This is because, since  $\mathcal{O}_{\partial a g}$ generates the $\mathcal{O}_{bsa}$ Dirac structure at one-loop. The resulting matching onto SCET then generates the same operator for the soft contribution, cf. Eq.~\eqref{eq:Obsa_A} and \eqref{eq:dirac_str}. The complete set of SCET operators generated in this work, from both $\mathcal{O}_{\partial ag}$ and $\mathcal{O}_{bsa}$, is collected in Table~\ref{tab:scet-basis} of Appendix~\ref{app:rg}, where their renormalization-group evolution is also discussed.

\section{Results}
\label{sec:sec6}
The complete $\zeta_K$ expression with $E_K$ (or equivalently, $m_a$) dependency is found in Ref.~\cite{Lu:2007sg} as
\begin{equation}\label{eq:fullzeta}
    \zeta_K(m_a)  = \frac{0.297}{1-1.28 (m_a^2/m_B^2)}\;.
\end{equation}
We use this expression in the numerical SCET results presented below.  Solely using $\zeta_K(q^2 = 0) = 0.297$ without $m_a$ dependence leads to prominent qualitative and quantitative differences between matrix element results from SCET and those found from the LCSR $f_0(q^2)$ form factor approach. Including the full $m_a$ dependence through Eq.~\eqref{eq:fullzeta}, much better agreement is reached. \\

\noindent\textbf{Case 1 -}
We first consider the case in which the Wilson coefficient of the operator $aG\widetilde G$ becomes nonzero only below the electroweak scale, i.e., at $\mu\sim m_b$. The numerical estimates obtained from Eqs.~\eqref{eq:spec}, \eqref{eq:soft} and using Table~\ref{tab:numbers} give
\begin{align}
R_{\rm spec}\equiv\left|\frac{\mathcal{M}_{\rm spec}^{\partial a g}}{\mathcal{M}_{\rm soft}^{\partial a g}}\right|\simeq 0.25-0.30\;,\label{eq:Rspec}
\end{align}
where the higher and lower value respectively corresponds to the chosen lower and upper bounds on the axion mass: $ 0.5$ GeV $\lesssim m_a  \lesssim 2.3$ GeV. The spectator-scattering contribution induced by \(\mathcal{O}_{\partial ag}\) is therefore somewhat smaller than the soft-overlap contribution, but it is neither parametrically nor numerically negligible. This result is consistent with the SCET factorization framework, in which soft overlap and hard spectator scattering both enter at leading order in the $1/m_b$ expansion~\cite{Bauer:2000yr,Beneke:2003pa}. Their relative size is consequently determined by the short-distance matching coefficients, the meson decay constants, the inverse moment $\lambda_B$, and the relevant light-cone distribution amplitudes, rather than by a suppression in the heavy-quark expansion.

Our calculation provides a quantitative realization of this structure for $B\to Ka$ in the low-energy gluonic axion theory. Starting from the dimension-seven operator $\mathcal{O}_{\partial ag}$, we calculate both the radiatively generated local-current contribution and the gluonic spectator-scattering contribution within a common SCET framework. The decay amplitude takes the schematic form
    \begin{align}
\mathcal{M}_{\rm tot}^{\partial a g}=\mathcal{M}_{\rm soft}^{\partial a g}+\mathcal{M}_{\rm spec}^{\partial a g}\;.
\end{align}
For constructive interference, the ratio above corresponds to
\begin{align}
\frac{\Gamma_{\rm tot}}{\Gamma_{\rm soft}}=\left|1+\frac{\mathcal{M}_{\rm spec}^{\partial a g}}{\mathcal{M}_{\rm soft}^{\partial a g}}\right|^2\simeq 1.5-1.7\;.
\end{align}
Thus, even though the spectator-scattering contribution is smaller than the soft-overlap term, it produces a sizeable correction to the total decay amplitude and should be retained in precision phenomenological analyses. Since the inferred sensitivity to the axion decay constant scales as $f_a^{\rm lim}\propto |\mathcal{M}_{\rm tot}|$, including the spectator term strengthens the corresponding limit on $f_a$ by approximately 25\%-30\%. The relative sign should, however, be retained explicitly, since destructive interference would instead reduce the total rate.

\begin{figure}[h]
    \centering
    \includegraphics[width=\linewidth, height=6.5cm]{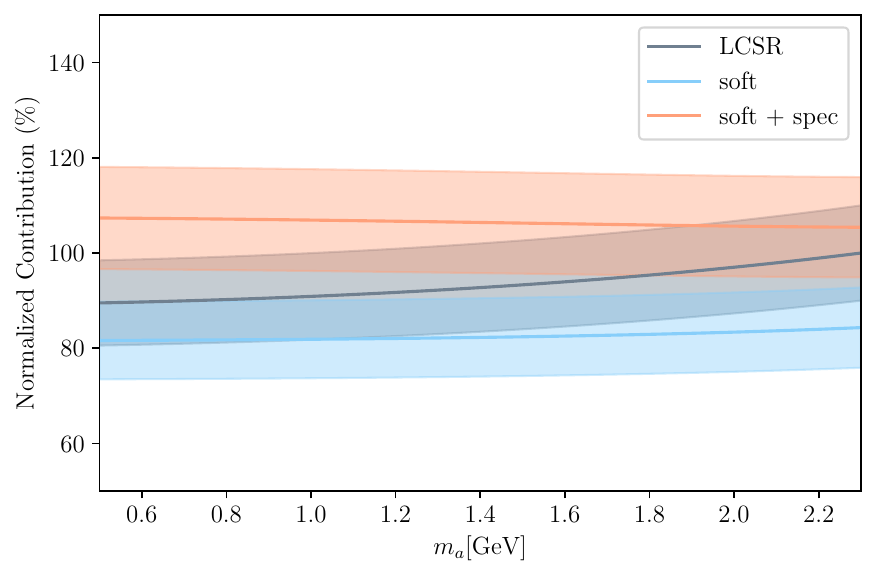}
    \caption{$\mathcal{O}_{\partial a g}$ mediated $B\to Ka$ matrix element results as a function of axion mass $m_a$. The gray, blue, and orange curves respectively denote the LCSR result, the soft-overlap SCET result, and the total soft plus spectator SCET result. The colored bands denote the uncertainties. See the discussion near Eq.~\eqref{eq:LCSR} for further details.
}
    \label{fig:plot}
\end{figure}
In Fig.~\ref{fig:plot} we show the comparison between the SCET results obtained in the present work and the usual approach~\cite{Bisht:2024hbs} of taking the QCD operator matrix element with the LCSR-derived form factor $f_0(q^2)$. With this approach, the matrix element is
\begin{equation}\label{eq:LCSR}
\begin{aligned}
 \mathcal{M}_{\text{LCSR}}^{\partial a g} &= \frac{\mathcal{C}_{\partial a g} g_s C_F }{16\pi^2 f_a} \bigg[\frac{17m_b^2 - 2m_a^2}{9} \bigg] \bigg(\frac{m_B^2 - m_K^2}{2}\bigg) f_0(m_a^2)\;,\\
 &\equiv \frac{\mathcal{C}_{\partial a g}g_s}{f_a} \times M^{\partial a g}_{\text{LCSR}} (m_a)\;.
\end{aligned}
\end{equation}
$M^{\partial a g}_{\text{LCSR}} (m_a)$ is an increasing function of $m_a$ and is plotted in Fig.~\ref{fig:plot} after normalizing by its maximum value at $m_a = 2.3\ \text{GeV}$. The SCET matrix elements are also plotted with $(\mathcal{C}_{\partial a g} g_s/f_a)$ extracted and the same normalization factor. The uncertainty in the LCSR and SCET soft-overlap results is approximately $10\%$ from $f_0(q^2)$ and $\zeta_K(m_a)$~\cite{Ball:2004rg, Ball:2004ye, Lu:2007sg}. The uncertainty in the spectator contribution arises largely from $\lambda_B$. Recent estimates~\cite{LatticeParton:2024zko} show $\lambda_B \simeq 0.31 - 0.44$ GeV implying an uncertainty of $20-30\%$ (see also Appendix~\ref{app:Numbers}). Roughly, this corresponds to $10 -15\%$ uncertainty in the total SCET result, and the plot shows the $10\%$ uncertainty band. The spectator contribution makes the SCET result larger than LCSR (still overlapping within the uncertainties), and makes it a \textit{decreasing} function of $m_a$.

To illustrate the maximal experimental sensitivity within the heavy QCD axion scenario, we combine the soft-overlap and spectator-scattering contributions in Eqs.~\eqref{eq:spec} and \eqref{eq:soft} to obtain
\begin{align}
    &\mathrm{Br}(B^{\pm}\to K^{\pm}a) \nonumber \\
    & \simeq 
     1.53\times10^{-8} \left(\frac{5\,{\rm GeV}}{f_a}\right)^2
    \left[1 -0.17\left(\frac{m_a}{2\,{\rm GeV}}\right)^2\right]\;.
\label{eq:bka1}
\end{align}
Since the effective interaction is assumed to be generated directly at the scale $\mu\sim m_b$, no large logarithms arise from renormalization-group evolution. Consequently, the resulting sensitivity to the axion decay constant is rather modest, probing only $f_a=\mathcal{O}({\rm GeV})$. Nevertheless, this reach remains competitive with existing bounds on hadronic axion-like particles from pion~\cite{Altmannshofer:2019yji} and kaon~\cite{Gori:2020xvq} decays.

The predicted branching fraction in Eq.~\eqref{eq:bka1} lies well below the current sensitivity of the $B$ factories. Nevertheless, it is instructive to estimate the ultimate reach of an idealized Belle II search assuming negligible background, as motivated by displaced vertex searches~\cite{Bertholet:2021hjl, Bandyopadhyay:2022klg}. For the full integrated luminosity of $50\,{\rm ab}^{-1}$, a zero-background analysis corresponds to an upper limit of approximately $2.3$ signal events at $90\%$ confidence level. The expected signal yield is
\begin{align}
N_{\rm sig}\simeq N_{B^\pm}\, {\rm Br}(B^\pm\to K^\pm a)\, {\rm Br}(a\to X_{\rm vis})\, \epsilon_{\rm rec}\;,
\end{align}
where $\epsilon_{\rm rec}$ denotes the reconstruction efficiency. The axion branching fractions can be computed systematically within chiral perturbation theory ($\chi$PT)~\cite{Bauer:2021wjo,Bandyopadhyay:2021wbb}, supplemented by vector-meson dominance~\cite{Fujiwara:1984mp} to account for resonance-mediated contributions~\cite{Aloni:2018vki,Cheng:2021kjg,Bisht:2024hbs}~\footnote{Also see embedding of axions in SMEFT~\cite{Galda:2021hbr, Grojean:2023tsd}, heavy hadron $\chi$PT~\cite{Mahajan:2022zer,Lu:2026hmc}, Wess-Zumino-Witten terms~\cite{Chakraborty:2024tyx,Bai:2024lpq}, and issues pertaining to regularization schemes involving axions~\cite{Quevillon:2021sfz,Bisht:2026qwe}.}. Although the absolute experimental reach is limited in this scenario, it is important to emphasize that this originates from the small short-distance production coefficient, and not from any suppression of the leading-power spectator-scattering contribution identified in this work.\\

\noindent\textbf{Case 2--} In this ultraviolet realization, the Wilson coefficient of the operator $aG\widetilde G$ is taken to be nonzero above the electroweak scale, generating a direct $b\to sa$ operator below the electroweak scale. In this case, the soft contribution comes directly from the operator rather than being one-loop generated and dominates over the spectator effect:
\begin{equation}
R_{\text{spec}}^{bsa}\equiv \left|\frac{\mathcal{M}_{\rm spec}^{bsa}}{\mathcal{M}_{\rm soft}^{bsa}}\right| \simeq 0.06~\text{--}~0.07\;.
\end{equation}
The hierarchy between soft overlap and spectator scattering therefore depends crucially on the scale at which the gluon-only boundary condition is imposed. Just like the previous case, the branching fractions can be estimated from Eq.\,\eqref{eq:spec1} and \eqref{eq:soft1} as 
\begin{align}
    \mathrm{Br}(B^{\pm}\,&\to K^{\pm}a)
    \simeq\,2.7~(17.3)\times10^{-6}\nonumber \\
    &\left(\frac{200\,\mathrm{GeV}}{f_a}\right)^{2}
    \left[1 - 0.05\left(\frac{m_a}{2\,\mathrm{GeV}}\right)^2\right]\;.
\end{align}
The numerical coefficients depend on the underlying ultraviolet completion. In the numerical analysis presented here, we assume the KSVZ (Kim-Shifman-Vainshtein-
Zakharov) model~\cite{Kim:1979if,Shifman:1979if} with UV scales of $1 (10)~\mathrm{TeV}$. The small contribution of spectator scattering to the total amplitude in this UV realization justifies the standard approach of taking the LCSR form factor with $\simeq 10\%$ uncertainty.

Finally, from the results of Section~\ref{sec:sec4}, we estimate the ratio of spectator and soft contributions coming from the chromomagnetic $\mathcal{O}_{8g}$ operator
\begin{equation}
    R_{\text{spec}}^{8g} \equiv \left|\frac{\mathcal{M}_{\rm spec}^{8 a g}}{\mathcal{M}_{\rm soft}^{8 g}}\right|\simeq 0.5 - 0.9 \;.
\end{equation}
So, although both $\mathcal{O}_{8g}$ and $\mathcal{O}_{\partial a g}$ give rise to leading spectator and soft effects in a similar manner, the full calculations show that the spectator-to-soft ratio is numerically larger for $\mathcal{O}_{8g}$. 

It is also of importance to comment on the relative contributions of \text{different} operators to the matrix elements. As discussed previously, the Wilson coefficient $\mathcal{C}_{8g}$ is numerically small compared to $\mathcal{C}_{Dg}$. As a result, we find the contribution of $\mathcal{O}_{8g}$ to be smaller than $\mathcal{O}_{\partial ag}$. For e.g, from Eq.~\eqref{eq:8agspec} and \eqref{eq:spec},
\begin{equation}
   \bigg|\frac{\mathcal{M}_{\rm spec}^{8ag}}{\mathcal{M}_{\rm spec}^{\partial ag}} \bigg|\simeq 0.15\;.
\end{equation}
On the other hand, in the second UV realization, the direct operator $\mathcal{O}_{bsa}$ contribution dominates the decay amplitude, yielding substantially stronger limits on $f_a$~\cite{Bisht:2024hbs}, while the gluonic spectator term becomes a comparatively small correction. For $\mathcal{C}_{bsa} = 2.46\times 10^{-7}$~\cite{Bisht:2024hbs}, the ratio of spectator contributions from $\mathcal{O}_{\partial a g}$ and $\mathcal{O}_{bsa}$ is very small: 
\begin{equation}
    \bigg|\frac{\mathcal{M}^{\partial a g}_{\text{spec}}}{\mathcal{M}^{bsa}_{\text{spec}}}\bigg| \simeq  0.02\;. 
\end{equation}
To conclude, in the first UV realization, both the local $b\to sa$ interaction and the spectator amplitude originate from $\mathcal{O}_{\partial ag}$, and the two contributions are naturally comparable. Nevertheless, their common short-distance coefficient is strongly suppressed by the flavor-changing weak transition. Consequently, the large relative spectator correction does not translate into a strong absolute constraint on $f_a$. This situation should be contrasted with the second UV scenario considered in Refs.~\cite{Chakraborty:2021wda,Bisht:2024hbs}, where the operator $aG\widetilde G$ is defined above the electroweak scale. In that case, electroweak matching together with renormalization-group evolution generates additional flavor-changing operators, leading to a substantially enhanced production rate and much stronger constraints on $f_a$. \\

{\bf Future Directions:} Several extensions of this framework appear promising. A natural next step is to generalize the present analysis by incorporating a complete set of axion interactions, including the axion-quark couplings. In addition, since $\mathcal{O}_{\partial ag}$ carries an explicit gluon at tree level, this hard-collinear gluon can result in a multi-hadron collinear state such as in $B\to K \pi a$. Instead of a single $\phi_K(u)$, here a two-meson distribution amplitude $\Phi_{K\pi}(u,\zeta,m_{K\pi}^2)$ appears. This topology has no counterpart for the direct current operator $\mathcal{O}_{bsa}$. In this three-body decay, the Dalitz distribution can resolve the soft vs spectator contributions. Moreover, a distinction can be made whether the final state $K\pi$ appears as a resonant or continuum structure. The framework extends further by treating the axion itself as a collinear field when $m_a \sim m_K$. In that regime $aG\widetilde{G}$ acts within the collinear sector rather than combining into the hard vertex. This opens up interesting kinematic configurations for which our SCET treatment of the axion is most suitable. Furthermore, a combined analysis of axion involved modes such as $B\to K^{(*)}a$, $B\to K\pi a$, $B\to\pi a$, $B_s\to\phi a$ and $\Lambda_b\to\Lambda a$ within the SCET framework, resolving the soft-overlap and spectator contributions, would provide an indirect handle on $C_{gg}/f_a$ and on the axion mass. A combined SCET treatment at the partonic level together with symmetry-based mesonic effective theories for multi-hadron final states~\cite{Chakraborty:2026vdz, Chakrabortty:2026hilbert} may provide a systematic framework for studying exclusive heavy-axion decays beyond two-body modes.

\begin{acknowledgments}
The authors thank Upalaparna Banerjee for correspondence and helpful discussions. We also thank Matthias Neubert for valuable comments and feedback on the manuscript. S.C. thanks the IIT-Kanpur initiation grant (PHY/2022220) and the Science and Engineering Research Board, Government of India (Grant No. SRG/2023/001162) for financial support. S.C. and A.S. gratefully acknowledge the hospitality of the Mainz Institute for Theoretical Physics (MITP), Germany, during the workshop ``ALPs Across Scales", and S.K.\ acknowledges the organizers and participants of the workshop ``Higgs and Effective Field Theory (HEFT 2026)", Valencia, Spain, where part of this work was carried out. 
\end{acknowledgments}

\appendix
\section{Numerical parameters}\label{app:Numbers}
In Table~\ref{tab:numbers}, we collect the numerical values of the parameters involved in our estimates. We take $\lambda_B = 338 \pm 68$ MeV from Ref.~\cite{Mandal:2023lhp}. Strong coupling constant $\alpha_s(5\ \text{GeV})$ is taken from~\cite{ParticleDataGroup:2024cfk}. Other parameter values not mentioned in the text are provided in PDG 2024~\cite{ParticleDataGroup:2024cfk}.
\begin{table}[h]
\centering
\renewcommand{\arraystretch}{1.4}
  \begin{tabular}{|l l|| l l|}
    \hline
    \textbf{Parameter} & \textbf{Value (GeV)} & \textbf{Parameter} & \textbf{Value (GeV)} \\
    \hline
    \hline
    $m_b$ & $4.18$  & $\tau_B$ &  $2.4 \times 10^{12}$ \\
    $m_B$ & $5.28$  & $E_0$ &  $5$ \\
    $m_K$ & $0.49$  & $E_0'$ &  $0.15$ \\
    $m_u\ (2\ \text{GeV})$ & $0.003$  & $m_d\ (2\ \text{GeV})$ &  $0.007$ \\
    $m_{\pi^0}$ & $0.135$  & $m_s\ (2\ \text{GeV})$ &  $0.12$ \\
     $m_{\eta}$ & $0.548$  & $m_{\eta'}$ &  $0.958$ \\
      $f_\pi$ & $0.13$  & $\text{Br}(B\to K \pi^0)$ &  $1.3\times 10^{-5}$ \\
       $\text{Br}(B\to K \eta)$ & $1.2\times 10^{-6}$  & $\text{Br}(B\to K \eta')$ &  $6.3\times 10^{-5}$ \\
    $f_B$ & $0.19$  & $\lambda_B$ &  $0.338$ \\
    $f_K$ & $0.156$ &  $\lambda_t = V_{ts}V_{tb}$  & $0.042$\\
    $G_F$  &  $1.166\times 10^{-5}$  &$\alpha_s(m_b)$ & $0.22$ \\
    $C_F$ & $4/3$ & $N_c$ & $3$\\
   \hline\hline
  \end{tabular}
  \caption{Numerical parameters used in phenomenology. Values for quantities with mass dimensions are in appropriate powers of GeV.}
  \label{tab:numbers}
\end{table}

\section{SM decays and ALP mixing}\label{app:mixing}

A comparison of our $B\to Ka$ analysis with SCET studies of $B$ decays to SM final states including isosinglet mesons~\cite{Williamson:2006hb} such as $\eta'$ is worth noting. The flavor-singlet axial vector current is anomalous in QCD, which leads to the $G\widetilde{G}$ coupling of flavor-singlet mesons. In the low-energy Chiral Lagrangian, this anomalous coupling is shown to contribute to the gluonic mass term~\cite{Bass:2018xmz}. However, in $B$ decays occurring at the hard $m_B$ scale, the $\eta'$ is treated at the parton level, and the relevant operator is not the Chiral Lagrangian anomalous coupling. Hence, the gluon equation of motion to eliminate $\mathcal{O}_{Dg}$ does not generate an operator analogous to $\mathcal{O}_{\partial a g}$ for isosinglet mesons. Still the possibility of generating isosinglet quark current producing $\eta'$ from two gluons is present, mediated by $\text{SCET}_{\text{I}}$ operators $Q_{gs}^{(0)}, Q_{gs}^{(1)}$ given in Ref.~\cite{Williamson:2006hb}. However, being subleading in the $\alpha_s$ expansion compared to the pure four-quark $\text{SCET}_{\text{I}}$ operator contributions, these are not calculated. The latter descend from WET four-fermi and penguin operators containing isosinglet quark currents, and give rise to the leading spectator and soft-overlap contributions. Thus, the SCET-level analysis in this case is qualitatively different from the present work, where the $aG\tilde{G}$ coupling is present in the QCD Lagrangian itself at the $m_b$ scale.

From a phenomenological perspective, $B \to KP$ decays with pseudoscalar mesons $P = \{ \pi^0, \eta, \eta' \}$ become relevant to the present work because of the mesons mixing with axions. Apart from the direct operator-mediated production of axions $B\to Ka$, as obtained in this work, the axion can be produced via mixing with the final-state mesons in $B\to KP$. The ALP-meson mixing has been well studied in the literature within the framework of the chiral Lagrangian~\cite{Georgi:1986df, Aloni:2018vki, Bauer:2021wjo}. It arises from chiral rotation of the quark fields to absorb the $aG\widetilde G$ coupling into axion-quark couplings. This implies effective ALP-quark couplings $\hat{\mathcal{C}}_{qq} = \mathcal{C}_{qq} + c_{gg} \kappa_q$ where $q = \{u,d,s \}$. Within our framework, $\mathcal{C}_{qq} = 0$ and $\mathcal{C}_{gg} =  c_{gg}\ (\alpha_s/4\pi ) $. The mixing angle between the axion and the pseudoscalar $P$ contains kinetic and mass-mixing terms. Taking the conventional choice of $\bm{\kappa}_q = \bm{m}_q^{-1}/{\rm Tr}(\bm{m}_q^{-1})$ eliminates the mass-mixing term and the mixing angle is then~\cite{Aloni:2018vki}
\begin{equation}
    \theta_{aP} = \frac{f_\pi}{2\sqrt{2} f_a} \frac{m_a^2 K_{aP}}{m_P^2 - m_a^2}\;,
\end{equation}
where $K_{aP}$ for the three mesons is given by
\begin{equation}
    \begin{aligned}
        K_{a\pi^0} &= c_{gg}(\kappa_u - \kappa_d)\;, \\
        K_{a\eta} &= \sqrt{\frac{2}{3}}c_{gg}(\kappa_u + \kappa_d - \kappa_s)\;, \\
        K_{a\eta'} &=  \sqrt{\frac{1}{3}} c_{gg}(\kappa_u + \kappa_d + 2\kappa_s)\;.
    \end{aligned}
\end{equation}
Thus, the total production of axions in $B$ decays to kaons receives contributions from both direct and indirect mixing-induced effects. At the amplitude level,
\begin{equation}
\begin{aligned}
    \mathcal{M}(B\to Ka) &= \mathcal{M}_{\text{direct}} + \mathcal{M}_{\text{mixing}}\;, \\
   \mathcal{M}_{\text{mixing}} &= \sum_{P} \mathcal{M}(B\to KP)\theta_{aP}\;.
\end{aligned}
\end{equation}
 Therefore, three terms contribute to the branching fraction. The first term is the direct contribution $\propto |\mathcal{M}_{\text{direct}}|^2$. In the absence of mixing, this is the only contribution, and our numerical results in Section~\ref{sec:sec6} are based on it. The other two terms arise due to non-zero mixing with mesons and are proportional to $ |\mathcal{M}_{\text{mixing}}|^2$ and $|\mathcal{M}_{\text{direct}}||\mathcal{M}_{\text{mixing}}| $. We call them the pure mixing term and the interference term, respectively.
To get a rough estimate of the size of these terms, we are ignoring phase or sign effects in the discussion. Then only the branching fractions of $B\to KP$ (inferred experimentally) and the mixing angles given above are needed. 

For Case 2, owing to typical $f_a \simeq 200$ GeV $\gg f_\pi$, the mixing angles are very small. Therefore, both the pure mixing and interference terms in the branching fraction are negligible compared to the direct term. On the other hand, Case 1 mediated by $\mathcal{O}_{\partial a g}$ has typical $f_a \simeq 5$ GeV. Due to $m_a > 0.5$ GeV, the pion mixing contribution is negligible. We can also ignore $\eta$-mixing since $\text{Br}(B\to K\eta) \ll \text{Br}(B\to K\eta')$ (see Table~\ref{tab:numbers}). Considering only the $\eta'$ contribution, we find that away from resonance (i.e. a few hundred MeVs beyond $m_{\eta'}$), the pure mixing term remains $\lesssim 10\%$ of the direct term. On the other hand, the interference term can be as large as $\simeq 50\%$ of the direct term. A detailed investigation, including the phases, therefore seems phenomenologically relevant for Case 1 but lies outside the main scope of this work.

\section{Comment on the RGEs of the axion involved SCET operators}\label{app:rg}
\begin{table*}[t]
	\centering
	\renewcommand{\arraystretch}{1.6}
	\setlength{\tabcolsep}{6pt}
	\begin{tabular}{|l| l | c|}
		\hline
		\multicolumn{2}{|c|}{Axion-involved aSCET operator} & \multicolumn{1}{|c|}{Corresponding SCET current in SM} \\
		\hline
        \hline
		$\mathcal{O}^{\rm SCET}_{\partial a g}$ &
		$\big[(\bar\xi_n W_n)\,\gamma_\perp^{\mu}\,P_L\,T^A\,h_v\big]\,
		\epsilon^{\perp}_{\mu\sigma}\,(n\!\cdot\!\partial a)\,(\bar n\!\cdot\!\partial)\,
		\mathcal{B}^{A\sigma}_{n\perp}$ &
		$J^{B}_{S}$\\
		
		
		\hline
		
		$\mathcal{O}^{(B)\,\rm SCET}_{bsa\,1}$ &
		$(n\!\cdot\!\partial a)\,(\bar\xi_n W_n)\,P_R\,\slashed{\mathcal{B}}_{n\perp}\,h_v$ &
		$J_S^B$\\
		
		$\mathcal{O}^{(B)\,\rm SCET}_{bsa\,2}$ &
		$(\bar n\!\cdot\!\partial a)\,(\bar\xi_n W_n)\,P_R\,\slashed{\mathcal{B}}_{n\perp}\,h_v$ & $J_S^B$ \\
		
		$\mathcal{O}^{(B)\,\rm SCET}_{bsa\,3}$ &
		$(\partial_\mu a)\,(\bar\xi_n W_n)\,P_R\,\slashed{\mathcal{B}}_{n\perp}\,\gamma_\perp^{\mu}\,h_v$ &
		$J_{V_1}^B$ \\
		
		\hline
		$\mathcal{O}^{\rm SCET}_{8 a g}$ &
		$a\,\big(\bar\chi_n\,\gamma_\perp^{\mu}\,P_L\,T^A\,Y_s^{\dagger}\,h_v\big)\,
		\epsilon^{\perp}_{\mu\nu}\,(\bar n\!\cdot\!\partial)\,\mathcal{B}^{A\nu}_{n\perp}$ &
		$J^{B}_{S}$ \\
        \hline
		
		$\mathcal{O}^{(A)\,\rm SCET}_{bsa\,1}$ &
		$(n\!\cdot\!\partial a)\,(\bar\xi_n W_n)\,P_R\,h_v$ &
		$J^{A}_{S}$
		 \\
		
		$\mathcal{O}^{(A)\,\rm SCET}_{bsa\,2}$ &
		$(\partial_\mu a)\,(\bar\xi_n W_n)\,\gamma_\perp^{\mu}\,h_v$ &
		$J^{A}_{V_1}$\\
		
		$\mathcal{O}^{(A)\,\rm SCET}_{bsa\,3}$ &
		$\epsilon_\perp^{\mu\nu}\,(\partial_\mu a)\,(\bar\xi_n W_n)\,\gamma^{\perp}_{\nu}\,h_v$ &
		$J^{A}_{V_1}$ \\
		\hline
        ${\cal O}_{agg}^{\rm SCET}$ & $a \,\epsilon^{\perp}_{\mu\nu}\,(\bar n \cdot \partial {\cal B}_{n\perp}^{\mu}) (n \cdot \partial {\cal B}_{\bar n\perp}^\nu)$ & NA\\
		\hline
	\end{tabular}
	\caption{SCET operator basis in the axion-extended SM (left) and the corresponding SM heavy-to-light SCET currents of \cite{Hill:2004if}(right).}
	\label{tab:scet-basis}
\end{table*}

The axion-involved SCET operators (`aSCET'), generated in our analysis, are collected in Table~\ref{tab:scet-basis}. This is not an exhaustive list. These are the operators generated at the leading order with the minimal gluonic scenario and contributing to the decay mode $B\to K a$. The operator normalization follows Eqs.\,\eqref{eq:Oag-operator}, \eqref{eq:Obsa_B}, \eqref{eq:Obsa_A} and \eqref{eq:O8g} with the soft Wilson lines kept un-decoupled. Throughout this appendix, we briefly discuss the RG below the hard scale $(\mu_h\sim m_B)$. Comments on the RG running above this scale are mentioned in Sec.~\ref{sec:sec2}.

For each operator, the third column records the SM heavy-to-light SCET current of Ref.~\cite{Hill:2004if} that remains after the axion field is extracted. The SCET currents as given in \cite{Hill:2004if} at the leading order are
\begin{align}
J^{A}_{S} &= (\bar \xi_n W_n) \;h_v\;,\\
J^{B}_{S} &= (\bar \xi_n W_n)\,
\slashed{\mathcal{B}}_{n\perp}\,h_v\;,\\
J^{B\,\mu}_{V1} &= (\bar \xi_n W_n)\,\slashed{\mathcal{B}}_{n\perp}(r\bar n)\,\gamma^{\mu}\,h_v\;.
\end{align}

The axion is a singlet under all SM gauge symmetries, and for an on-shell external axion the pre-factors $n\!\cdot\!\partial a$, $\bar n\!\cdot\!\partial a$ and $a$ reduce to complex numbers. In $\mathcal{O}_{\partial ag}^{\rm SCET}$ and $\mathcal{O}_{8ag}^{\rm SCET}$ the transverse tensor is removed using 
\begin{equation}
    \epsilon^{\perp}_{\mu\sigma}\gamma^{\mu}_{\perp}P_L =\mp i\,\gamma^{\perp}_{\sigma}P_L\;,
\end{equation} 
after which both reduce to the scalar $B$-type structure $\bar\chi_n\slashed{\mathcal{B}}_{n\perp}P_L h_v$. Since $\bar\chi_n\gamma^{\mu}_{\perp}P_L=\bar\chi_n P_R\gamma^{\mu}_{\perp}$, this is the same chiral structure that appears in $\mathcal{O}^{(B)}_{bsa\,1,2}$. The QCD color and Dirac content of all four operators is therefore identical to $J^B_S$, and the axion enters only as an external singlet field. 

The operator ${\cal O}_{agg}^{\rm SCET}$, which has no SM analogue. This operator ${\cal O}_{agg}^{\rm SCET}$ does not contribute to any diagram of $B\to K a $ decay mode; however, we kept it in the table for completeness and to emphasize its potential role in generic $m_a$ scenarios. Moreover, ${\cal O}_{agg}^{\rm SCET}$ allows additional diagrams that could, in principle, contribute to the anomalous dimensions of the operators listed in Table~\ref{tab:scet-basis}. We stress that in the kinematic region considered here, with $m_a \sim 2~\mathrm{GeV}$, the axion is a heavy degree of freedom. Thus, one of the gluons attaching to the axion in the loop carries hard momentum which gets integrated out at the hard scale. As a result, within the axion-extended SCET framework considered here, the axion couples only at the hard vertex~\footnote{We note that for $m_a^2\lesssim m_b\Lambda_{\rm QCD}$ the gluon pair produced at the vertex may instead be hard-collinear; the matching is then not confined to the hard scale and genuinely new SCET$_{\rm I}$ structures can arise. This regime lies outside the mass window considered in this section}. With no new one-loop topologies contributing to the anomalous dimension, the SM results of Ref.~\cite{Bauer:2000yr, Hill:2004if} therefore apply directly to the operator list in aSCET mentioned in Table~\ref{tab:scet-basis}.

In the SM sector, the $A$-type and $B$-type currents do not mix. Correspondingly, the $A$-type and the $B$-type aSCET operators do not mix under RG running. The $A$-type operators do not mix among themselves, since time-ordered products with the SCET$_{\rm I}$ Lagrangian cannot connect different Dirac structures, and all $A$-type currents share the universal anomalous dimension~\cite{Hill:2004if}
\begin{equation}
    \gamma^A=-\Gamma_{\rm cusp}\ln(\mu/2E)+\tilde\gamma\;. 
\end{equation}

Note that $\mathcal{O}^{(A)}_{bsa\,3}$ and $\mathcal{O}^{(A)}_{bsa\,2}$ both, like $\mathcal{O}^{(B)}_{bsa\,3}$, involve $\partial_\perp a\sim\lambda^2$ and thus are subleading and dropped in the analysis of Sec.~\ref{sec:sec4}. The surviving $B$-type sector relevant here contains only the scalar structure which is multiplicatively renormalized. Because the hard-collinear gluon carries a fraction of the large momentum, the evolution is a convolution,
\begin{align}
	\frac{d}{d\ln\mu}\,\mathcal{C}^{B}(E,u,\mu) &=  \int_0^1\!dv\;\gamma^{B}(v,u;\mu)\,\mathcal{C}^{B}(E,v,\mu)\;,
\end{align}
with a kernel that separates into cusp, local non-cusp, and momentum-transfer pieces~\cite{Hill:2004if},
\begin{align}
	\gamma^{B}_S(u,v;\mu)&=\delta(u-v)\Big[-\Gamma_{\rm cusp}(\alpha_s)\ln\frac{\mu}{2E}
	\nonumber\\
    &~~+\tilde\gamma_B(\alpha_s,u)\Big]+u\,\mathcal{V}(u,v)\;.
	\label{eq:gammaB}
\end{align}

The only structural difference relative to the SM case is the extra $\bar n\cdot\partial$ acting on $\mathcal{B}_{n\perp}$ in $\mathcal{O}_{\partial ag}$, which in momentum-fraction space is a weighting by the collinear fraction $u$. Then the object obeying the standard evolution equation is $u\,{\cal C}^{\rm SCET}_{\partial ag}(E,u)$ rather than ${\cal C}_{\partial ag}(E,u)$. 

Since $\Gamma_{\rm cusp}$ enters $\gamma^A$ and $\gamma^B_S$ with the same coefficient, the leading logarithms cancel in the ratio $R_{\rm spec}=|\cal M_{\rm spec}/\cal M_{\rm soft}|$ of Eqs.\,\eqref{eq:Rspec}, leaving the difference of non-cusp terms and the $u$-dependent evolution of the B-type coefficient. For a pseudoscalar final state, the latter falls, where the $u$-dependent distortion is known to be small \cite{Hill:2004if}. We therefore expect resummation effects at the level of $10\%$--$20\%$ on the individual amplitudes, with partial cancellation in $R_{\rm spec}$. Below $\mu_i\sim\sqrt{2E\Lambda_{\rm QCD}}$, the spectator term factorizes into $\phi_B\otimes J\otimes\phi_K$, and the RG evolution is that of the LCDAs and would contribute to the uncertainties, for example, to $\lambda_B$ as mentioned in Table~\ref{tab:numbers}. 

A complete resummed analysis, particularly for generic axion masses, lies beyond the scope of the present work. Our primary objective here is to formulate the SCET framework for axion-induced heavy-to-light decays, leaving a systematic resummation and a comprehensive phenomenological analysis for future investigations.

\bibliography{SCET}

\end{document}